\documentclass[longauth]{aa} 
\usepackage[usenames,dvipsnames]{xcolor}

\usepackage{graphicx}
\usepackage{amsmath, bm}
\usepackage{natbib}
\bibpunct{(}{)}{;}{a}{}{,}

\usepackage{tikz}
\usepackage{txfonts}
\usepackage{longtable}
\usepackage{listings}
\usepackage{color}
\usepackage{textcomp}
\usepackage{multirow}
\usepackage{tablefootnote}
\usepackage{soul}
\usepackage{hyperref}

\defcitealias{Queiroz2023}{Q23}

\begin{document}

   \title{Transferring spectroscopic stellar labels\\ to 217 million {\it Gaia} DR3 XP stars with {\tt SHBoost}
   \thanks{Table \ref{tab:datamodel1} is available in electronic form at \url{data.aip.de/projects/shboost2024.html} and at the CDS via anonymous ftp to \url{cdsarc.u-strasbg.fr} (130.79.128.5) or via \url{http://cdsweb.u-strasbg.fr/cgi-bin/qcat?J/A+A/}}}
   \author{A. Khalatyan\inst{1}
          \and F. Anders\inst{2,3,4}\thanks{Corresponding author: \url{fanders@icc.ub.edu}}
          \and C. Chiappini\inst{1}
          \and A. B. A. Queiroz\inst{5,6}
          \and S. Nepal\inst{1,7}
          \and M. dal Ponte\inst{8}, \\ 
          C. Jordi\inst{4}
          \and G. Guiglion\inst{9, 10, 1}
          \and M. Valentini\inst{1}
          \and G. Torralba Elipe\inst{11, 12, 13} 
          \and M. Steinmetz\inst{1}
          \and M. Pantaleoni-González\inst{14,15},\\
          S. Malhotra\inst{2,3,4}
          \and Ó. Jiménez-Arranz\inst{16,2,3,4}
          \and H. Enke\inst{1}
          \and L. Casamiquela\inst{17}
          \and J. Ardèvol\inst{2,3,4} 
          }
   \institute{{Leibniz-Institut f\"ur Astrophysik Potsdam (AIP), An der Sternwarte 16, 14482 Potsdam, Germany}
        \and{Departament de Física Quàntica i Astrofísica (FQA), Universitat de Barcelona, C Martí i Franqués, 1, 08028 Barcelona, Spain}
        \and{Institut de Ciències del Cosmos (ICCUB), Universitat de Barcelona (UB), C Martí i Franqués, 1, 08028 Barcelona, Spain}
        \and{Institut d'Estudis Espacials de Catalunya (IEEC), Edifici RDIT, Campus UPC, 08860 Castelldefels (Barcelona), Spain}
        \and{Instituto de Astrof\'isica de Canarias, E-38200 La Laguna, Tenerife, Spain}
        \and{Departamento de Astrof\'isica, Universidad de La Laguna, E-38205 La Laguna, Tenerife, Spain}
        \and{Institut f\"{u}r Physik und Astronomie, Universit\"{a}t Potsdam, Haus 28 Karl-Liebknecht-Str. 24/25, D-14476 Golm, Germany}
        \and{INAF - Osservatorio Astronomico di Padova, Vicolo dell’Osservatorio 5, I-35122, Padova, Italy}
        \and{Zentrum für Astronomie der Universität Heidelberg, Landessternwarte, Königstuhl 12, 69117 Heidelberg, Germany}
        \and{Max Planck Institute for Astronomy, K\"onigstuhl 17, 69117, Heidelberg, Germany}
        \and{CIGUS CITIC – Department of Computer Science and Information Technologies, University of A Coruña, Campus de Elviña s/n, A Coruña, 15071 Spain}
        \and{Escuela Superior de Ingeniería y Tecnología, Universidad Internacional de la Rioja, Spain}
        \and{Escuela de Arquitectura y Politécnica, Universidad Europea de Valencia, Spain}
        \and{Departamento de Astrofísica, Centro de Astrobiología (CSIC-INTA), Camino Bajo del Castillo s/n., 28692 Villanueva de la Cañada, Madrid, Spain}
        \and{Department of Astrophysics, University of Vienna, Türkenschanzstra\ss e 17, 1180 Wien, Austria}
        \and{Lund Observatory, Division of Astrophysics, Department of Physics, Lund University, Box 43, 22100 Lund, Sweden}
        \and{GEPI, Observatoire de Paris, Université PSL, CNRS, 5 Place Jules Janssen, 92190 Meudon, France}
         }
 
   \date{Received \today; accepted ...}
  
  \abstract{With {\it Gaia} Data Release 3 (DR3), new and improved astrometric, photometric, and spectroscopic measurements for 1.8 billion stars have become available. Alongside this wealth of new data, however, there are challenges in finding  efficient and accurate computational methods  for their analysis. In this paper, we explore the feasibility of using machine learning regression as a method of extracting basic stellar parameters and line-of-sight extinctions from spectro-photometric data. To this end, we built a stable gradient-boosted random-forest regressor ({\tt xgboost}), trained on spectroscopic data, capable of producing output parameters with reliable uncertainties from {\it Gaia} DR3 data (most notably the low-resolution XP spectra), without ground-based spectroscopic observations. Using Shapley additive explanations, we interpret how the predictions for each star are influenced by each data feature. For the training and testing of the network, we used high-quality parameters obtained from the {\tt StarHorse} code for a sample of around eight million stars observed by major spectroscopic stellar surveys, complemented by curated samples of hot stars, very metal-poor stars, white dwarfs, and hot sub-dwarfs. The training data cover the whole sky, all Galactic components, and almost the full magnitude range of the {\it Gaia} DR3 XP sample of more than 217 million objects that also have reported parallaxes. We have achieved median uncertainties of 0.20 mag in V-band extinction, 0.01 dex in logarithmic effective temperature, 0.20 dex in surface gravity, 0.18 dex in metallicity, and $12\%$ in mass (over the full {\it Gaia} DR3 XP sample, with considerable variations in precision as a function of magnitude and stellar type). We succeeded in predicting competitive results based on {\it Gaia} DR3 XP spectra compared to classical isochrone or spectral-energy distribution fitting methods we employed in earlier works, especially for parameters $A_V$ and $T_{\rm eff}$, along with the metallicity values. Finally, we  showcase some potential applications of this new catalogue, including extinction maps, metallicity trends in the Milky Way, and extended maps of young massive stars, metal-poor stars, and metal-rich stars).
  }
  

  \maketitle

\section{Introduction}\label{sec:intro}

\begin{figure*}\centering
        \includegraphics[width=0.98\textwidth]{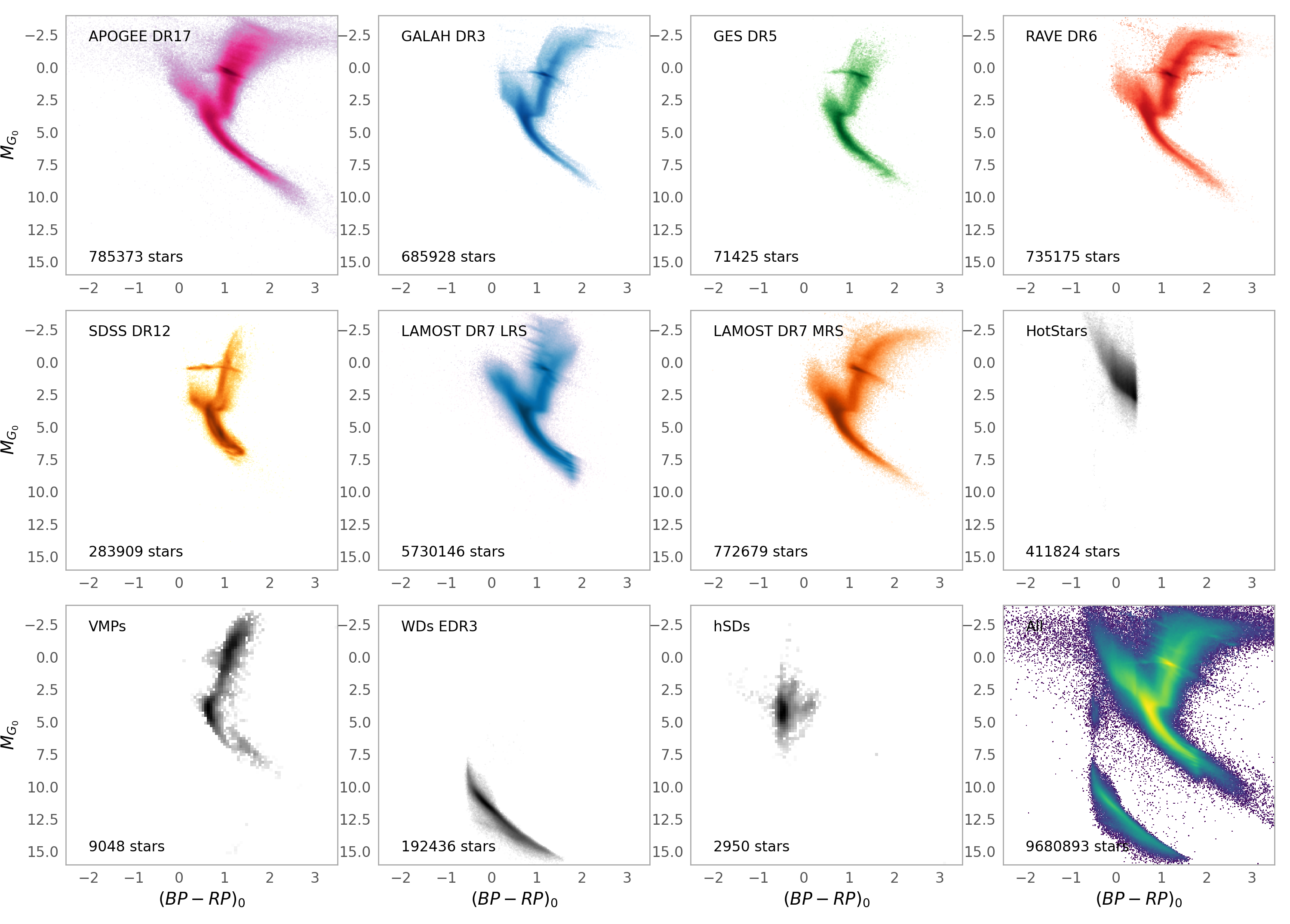}
        \caption{Distance- and extinction-corrected {\it Gaia} DR3 CMDs for each of the spectroscopic stellar surveys used in the training and test data for this work. The bottom right panel shows the joint dataset. Note:\ not all data points were used in the training and testing of the {\tt xgboost} models; duplicates were removed and different quality cuts applied for each training label. }
        \label{fig:trainingset_cmds}
\end{figure*}

Galactic Astronomy has become an exceedingly data-driven field with a vast and rapidly growing collection of multi-temporal and multi-wavelength data being observed by a multitude of stellar surveys. The {\it Gaia} mission \citep{GaiaCollaboration2016}, which is currently charting a six-dimensional (6D) phase-space map of the Milky Way, is currently providing accurate measurements for almost two billion stars in our Galaxy and the Local Group \citep{GaiaCollaboration2018, GaiaCollaboration2021, GaiaCollaboration2023V}. The volume of astronomical data available passed the 100 Terabyte ($10^{14}$ bytes) mark by the end 20th century and is expected to continue rapidly expanding towards the Petabyte ($10^{15}$ bytes) and Exabyte ($10^{18}$ bytes) scales in the near future \citep{Fluke2019, Vavilova2020}.

In response to the growing amount of data being collected, a variety of machine learning (ML) techniques have been integrated to complement more classical statistical methods of data analysis. These ML techniques are now at the forefront of data mining and analytics, thanks to their ability to cope with the current scale of the data and in anticipation of even greater volumes of data being released. Furthermore, the versatility of ML techniques allow us to use them in performing a wide variety of tasks, such as classification, regression, outlier detection, and data compression on large scales \citep[e.g.][]{Ivezic2014}. Overall, ML techniques have quickly become effective tools for astronomers to to turn a copious amount of raw data into useful knowledge \citep[for reviews, see e.g.][]{Baron2019, Fluke2019, Sen2022}. 

Recently, the scope of {\it Gaia} data has been expanded by the inclusion of more than 219 million low-resolution XP spectra \citep{Carrasco2021, DeAngeli2023, Montegriffo2023} in the third {\it Gaia} data release, DR3 \citep{GaiaCollaboration2023V}. For {\it Gaia} DR3, the mean XP spectra have primarily been used to infer much more precise stellar parameters, such as: $T_{\rm eff}, \log g, d, A_0,$ and [M/H] \citep{Andrae2023G, Fouesneau2023}. Various groups have already shown that this new dataset can potentially be used for a variety of Galactic archaeology purposes: to estimate rough stellar $\alpha$-element abundances \citep{Gavel2021, Hattori2024, Laroche2024, Li2024}, efficiently locate extremely metal-poor stars \citep{Witten2022, Xylakis-Dornbusch2022, Xylakis-Dornbusch2024, Lucey2023, Yao2024}, classify white dwarfs \citep{Echeverry2022, GaiaCollaboration2023M}, study hydrogen emission and absorption lines \citep{Weiler2023}, or improve the accuracy of stellar parameters determined from {\it Gaia} RVS spectra \citep{Guiglion2024}. 

Here, we aim to take advantage of the {\it Gaia} DR3 XP spectra by combining this dataset with additional information, both from {\it Gaia} and photometric databases. 
This paper aims to investigate the use of a relatively new but robust and well-tested regression technique, {\tt xgboost}, trained using spectroscopically derived data from a variety of stellar surveys, to determine the following stellar parameters: extinction, effective temperature, metallicity, surface gravity, and mass from astrometric and photometric data provided by the {\it Gaia} mission. Similarly to other ML techniques, {\tt xgboost}  achieves this by learning the complex, non-linear relationships between the raw input data and the desired outputs. 

For the training set, we mainly use the spectroscopy-derived data from \citet[][hereafter Q23]{Queiroz2023} obtained with {\tt StarHorse} \citep{Queiroz2018, Queiroz2020}, a Bayesian isochrone-fitting code that yields precise results for spectroscopic surveys. When only supplied with astrometric and multi-band photometric data, {\tt StarHorse} still produces very competitive results, compared to other techniques \citep[as we have shown in][]{Anders2019, Anders2022}. However, there are two main disadvantages with Bayesian isochrone fitting for these large datasets: 1) it is considerably slower and computationally more expensive, since the likelihood for each star has to be computed for a larger stellar-parameter range; and 2) it cannot make direct use of additional information (e.g. {\it Gaia} proper motions, colour-excess factor, or XP spectra\footnote{In principle, it is possible to use narrow- or intermediate-band photometry derived from the {\it Gaia} XP spectra to include their information content - however, this transformation adds additional systematic uncertainty, and in the case of {\tt StarHorse}, the execution time scales with the number of photometric observations included in the inference.}). Therefore, in this paper we explore the possibility of circumventing these problems with a well-trained ML algorithm, thus freeing computational resources and greatly improving the CO$_2$ budget of the new catalogue. In addition, by taking into account the information contained in the  {\it Gaia} DR3 XP spectra, we are able to improve, for instance, the inferred metallicities for a large sample of stars covering large portions of the Galaxy. We stress, however, that in this work we do not attempt to derive new age or distance estimates, since the information about these quantities is not directly encoded in the {\it Gaia} DR3 XP spectra.

This paper is structured as follows: in Sect. \ref{sec:data} we describe the data used to train and predict stellar parameters for {\it Gaia} stars. In Sect. \ref{sec:xgboost} we explain the concept and implementation of {\tt xgboost}, the supervised regression algorithm we use in this paper. Section \ref{sec:catalogue} presents the scope of the produced catalogues, including a discussion of the caveats of the present catalogue (Sect. \ref{sec:caveats}). Some immediate results derived from the catalogue are shown in Sect. \ref{sec:results}. We conclude the paper with a short discussion and our conclusions in Sect. \ref{sec:conclusions}. The appendices of the paper contain comparisons to other catalogues (App. \ref{sec:literature}),  data model of the produced catalogue (App. \ref{sec:datamodel}), and a possible re-calibration of the metallicity scale based on open and globular clusters (App. \ref{sec:feh_calib}).

\begin{table*}
    \caption{Cleaning conditions applied to each of the training labels.}
    \begin{tabular}{l|ccccc}
    Label & $A_V$ & $\log T_{\rm eff}$ & $\log g$ & [M/H] & Mass \\
    \hline
    Training range      & $[-0.2, 30]$ & $[3,5]$ & $[-1,9]$ & $[-4, 0.6]$ & $[0.05, 70]$   \\
    Maximum uncertainty & 0.1 mag or $0.1 \cdot A_V$ & $\sigma_{T_{\rm eff}}<0.1 \cdot T_{\rm eff}$ & 0.15 dex & 0.1 dex (0.2 dex for [M/H]$<-2$) & $0.2\cdot M$\\
    Excluded data       &     &      &   & WDs, hSDs & VMP\\
    \end{tabular}
    \label{tab:clean}
\end{table*}

\section{Data}\label{sec:data}

In supervised machine learning, the most important ingredient for building a well-performing classifier or regressor is the availability of a sufficiently large, precise, and accurate training dataset that covers as much of the parameter space as possible. For the case of the vast {\it Gaia} DR3 dataset (totalling more than 1.8 billion stars), a small but sizeable subset ($\gtrsim 7$ million stars) has already been co-observed spectroscopically by several major multiplex stellar surveys. This subset covers all major components of the Milky Way (discs, halo, bar, and bulge) and, in particular, the Apache Point Observatory Galactic Evolution Experiment  \citep[APOGEE;][]{Majewski2017} even exceeds the volume reached by {\it Gaia} towards the inner Galaxy. We can therefore consider the part of the {\it Gaia} DR3 catalogue observed by ground-based spectroscopic surveys a well-sampled subset of the full catalogue, at least down to the typical limiting magnitude of the most relevant spectroscopic surveys. 

\subsection{Spectroscopic labels}\label{sec:Q2023}

Our training set consists of roughly 7 million stars that have been observed by large ground-based spectroscopic stellar surveys, such as APOGEE,  GALactic Archaeology with HERMES survey (GALAH; \citealt{DeSilva2015}),  Gaia-ESO Survey \citep{Gilmore2012, Gilmore2022},  LAMOST surveys \citep{Cui2012, Deng2012},  RAdial Velocity Experiment (RAVE; \citealt{Steinmetz2006, Steinmetz2020}), and Sloan Extension for Galactic Understanding and Exploration (SEGUE; \citealt{Yanny2009}). For all these surveys, \citetalias{Queiroz2023}  reported  precise distances, extinctions, and stellar parameters using the Bayesian isochrone-fitting code {\tt StarHorse} \citep{Queiroz2018, Queiroz2020}. The data cover the main parts of the Hertzsprung-Russell diagram, from the hot main sequence (although sparsely sampled) to M-type stars (including dwarfs and asymptotic giant-branch stars).
For more information on the individual spectroscopic survey datasets, we refer to Sects. 3-4 and Figs. 1-5 of \citetalias{Queiroz2023}. 

To cover an important part the parameter range that has been missed by {\tt StarHorse} due to the lack of consistent stellar models in this regime, we added the HotPayne catalogue of LAMOST OBA stars \citep{Xiang2022}, a curated set of very metal-poor (VMP) stars\footnote{Primarily taken from the 2021 version of the SAGA database (\url{http://sagadatabase.jp/?p=43}; \citealt{Suda2008}), selecting Milky-Way stars observed with spectral resolution $R>5\,000$, and adding stars from \citet{Yong2021} and \citet{Li2022a}.}, along with the {\it Gaia} EDR3 white-dwarf catalogue of \citet{GentileFusillo2021} and the {\it Gaia} EDR3 hot sub-dwarf catalogue of \citet{Culpan2022} to this training set. 
The de-reddened and distance-corrected colour-magnitude diagrams (CMDs) of each of the training datasets are shown in Fig. \ref{fig:trainingset_cmds}. After joining the different datasets, we cleaned the sample from duplicate stars, giving preference to higher-resolution surveys and higher signal-to-noise (S/N) observations if multiple entries of the same stars were included.

\begin{figure*}\centering
        \includegraphics[width=0.99\textwidth]{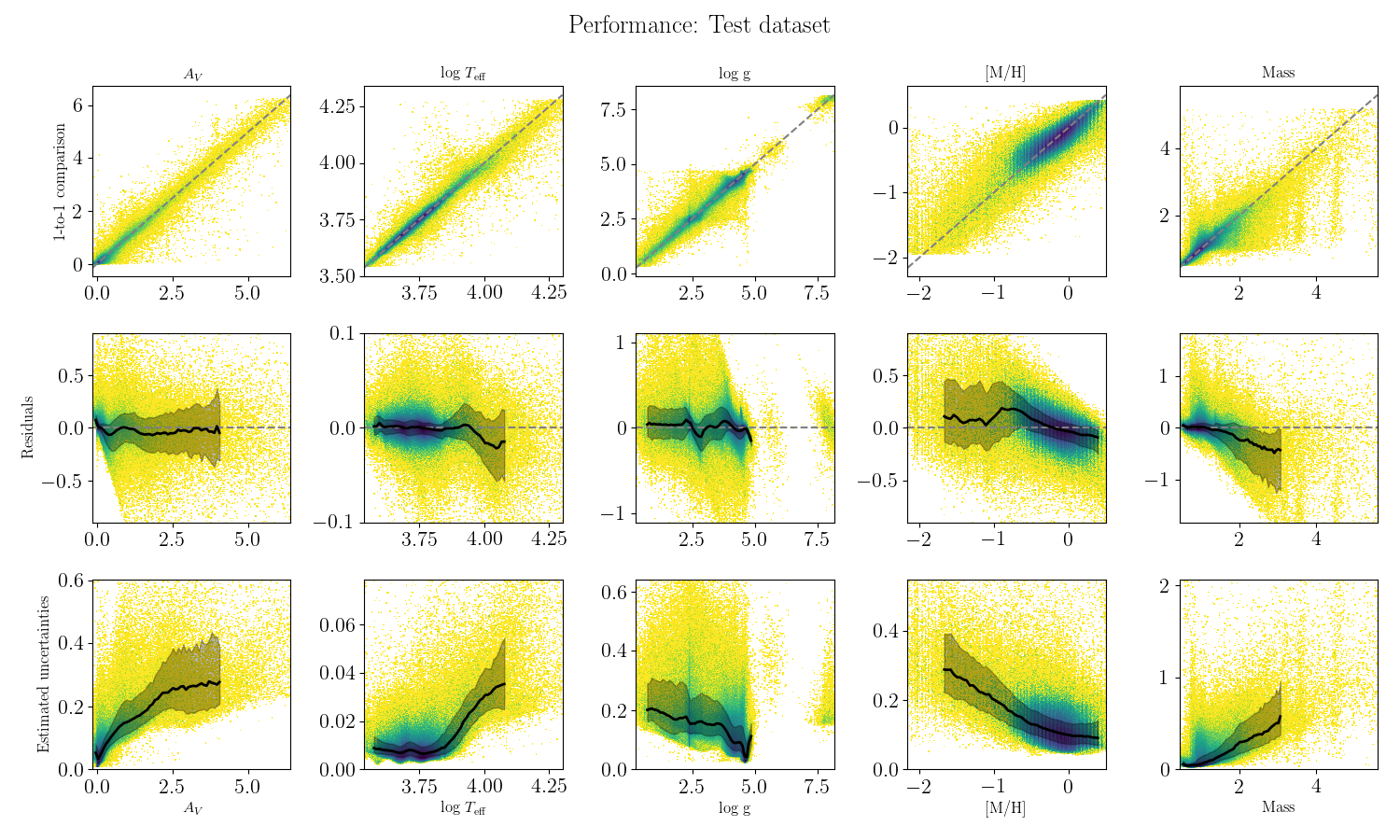}      
        \caption{Performance of the {\tt xgboost} models for the test datasets for each of the training labels. In the top row, we show the {\tt SHBoost} (mean) parameters predicted from {\it Gaia} DR3, 2MASS, and AllWISE against the spectroscopic values (test labels). The middle row shows the residuals (predicted:\ 'true'). The bottom row shows the formal uncertainties (derived with {\tt xgboost-distributions}). Each panel contains logarithmic density plots of the full sample of 217 million stars. The lines and shaded regions in the middle and bottom rows show the running median and 1$\sigma$ quantiles, respectively.
    }
        \label{fig:validation-onetoone}
\end{figure*}

\subsection{Training and test datasets}\label{trainingset}

We have aimed to keep the training data simple: we only used as our input columns (aka 'features') the {\it Gaia} DR3 XP spectral coefficients \citep{Carrasco2021, DeAngeli2023},  {\it Gaia} EDR3 astrometry ($l, b, \varpi, \sigma_{\varpi}, \mu_{ra}, \mu_{dec}$), and broad-band $\{G$, $BP$, $RP\}$ photometry \citep{GaiaCollaboration2021},  and {\it Gaia}  EDR3 astrometric fidelity flag {\tt fidelity\_v2} \citep{Rybizki2022}, as well as infrared (IR) $JHK_s$ and $W1W2W3W4$ magnitudes from 2MASS \citep{Cutri2003} and AllWISE \citep{Cutri2013}, respectively.
The XP coefficients have been normalised per star by the value of the first coefficient.

The ML algorithms (especially neural networks) are often susceptible to the varying quality of the training data. To test their impact, we performed tests using different quality cuts for the input data (i.e. varying the maximum uncertainty for each parameter; see second line in Table \ref{tab:clean}), finding no significant improvements. We therefore chose to use the full set of columns for each of the training datasets.

For each label (extinction $A_V$, effective temperature, $\log T_{\rm eff}$, surface gravity, $\log g$, metallicity, [M/H], and stellar mass, $M$), we applied a set of criteria to clean the training data from poorly determined or unreliable labels, as detailed in Table \ref{tab:clean}. For each label, 80\% of the data were used during the training phase (see the next section), while 20\% of the data are reserved for estimating the overall performance (test dataset; see Fig. \ref{fig:validation-onetoone}).  

\section{{\tt xgboost} regression}\label{sec:xgboost}

A large and still growing variety of ML algorithms exists on the market, each of them optimised for slightly different tasks. The problem we are dealing with here boils down to regression on tabular data. 
\citet{Borisov2021} and \citet{Grinsztajn2022} recently surveyed the performance of several state-of-the art ML regression techniques on tabular data. They concluded that algorithms based on gradient-boosted tree ensembles still outperform deep learning models for this kind of task (although some neural-network implementations, if sufficiently trained, can perform similarly well on large enough datasets; see e.g. \citealt{Klambauer2017}). Also, {\tt xgboost} has the additional advantage that it requires much less training time and parameter tuning \citep{Shwartz-Ziv2021}.

After some initial tests and complementary efforts using artificial neural networks, we therefore chose the best-performing tree-based algorithm considered by \citet[][see their Fig. 4]{Borisov2021}, {\tt xgboost}, for the purpose of this work. Following previous implementations (see \citealt{Anders2023Proc}), our approach was successfully adopted by other groups \citep{Rix2022, Andrae2023}. Overall,
{\tt xgboost} \citep{Chen2016} stands short for extreme gradient-boosted trees and it is a supervised ML algorithm that can perform both classification and regression. The algorithm has been used to some extent in astronomy for classification tasks \citep[e.g.][]{Bethapudi2018, Yi2019, Li2021, Cunha2022, Tolamatti2023, Xu2024}; more recently also its regression capability has been explored in some astronomical use cases -- e.g. photometric redshifts \citep{Li2022, Jia2023}, sunspot prediction \citep{Dang2022}, or spectroscopic stellar ages \citep{Hayden2022, He2022, Anders2023}.

In principle, there are various ways to estimate uncertainties with {\tt xgboost} (e.g. training the model on various subsets of the training data, or training models with a wide range of hyper-parameters). Perhaps the most rigorous way is to modify {\tt xgboost}'s cost function to include the prediction of confidence intervals in the inference \citep{Duan2019}. Thankfully, a fast implementation of this method based on the latest versions ($>2.0$) of the {\tt xgboost} library\footnote{\url{https://xgboost.readthedocs.io/}} is available in the {\tt python} module {\tt xgboost-distribution}\footnote{\url{https://xgboost-distribution.readthedocs.io/}}, which we therefore used to build our models. The results of this module can be interpreted as pseudo-posterior distributions (in the Bayesian terminology). As in \citet{Anders2023}, we provide both the 'traditional' {\tt xgboost} point estimates as well as the mean and standard deviation provided by {\tt xgboost-distribution}.

In addition to high accuracy, stability to noise, and relatively short training time, another advantage of using {\tt xgboost} is that it can be coupled with a powerful feature-attribution code, SHAP \citep{Lundberg2018}, which makes tree-based regression methods much more interpretable, even at the level of individual data points. In this manner, {\tt xgboost} is no longer an ML 'black box', but it also allows us to measure which of the input data (features) carry most of the information about the desirable parameters (labels). We analyse the SHAP values in more detail in Sect. \ref{sec:shap}.

\subsection{Tuning hyper-parameters}\label{sec:hyperparams}

Like most ML techniques, {\tt xgboost} has a number of hyper-parameters that can be tuned to optimise the performance. We tested each of the main parameters individually as well as with {\tt scikit-learn}'s {\tt GridSearchCV}, a cross-validation algorithm that can be used to loop over hyper-parameter configurations. Our tests confirmed that {\tt xgboost} is extremely stable under different configurations. The range of hyper-parameters that can be considered close to optimal is quite large. Using a sub-sample of 1\% of the training and test data, we found that for our case the following configuration yields optimal results in an acceptable runtime: {\tt learning\_rate}$=0.1-0.2$, {\tt max\_depth}$=6-10$, {\tt min\_child\_weight}$=1$, and {\tt subsample}$=0.8$.

\subsection{Typical accuracy and precision}\label{sec:errors}

\begin{figure*}\centering
        \includegraphics[width=1.0\textwidth]{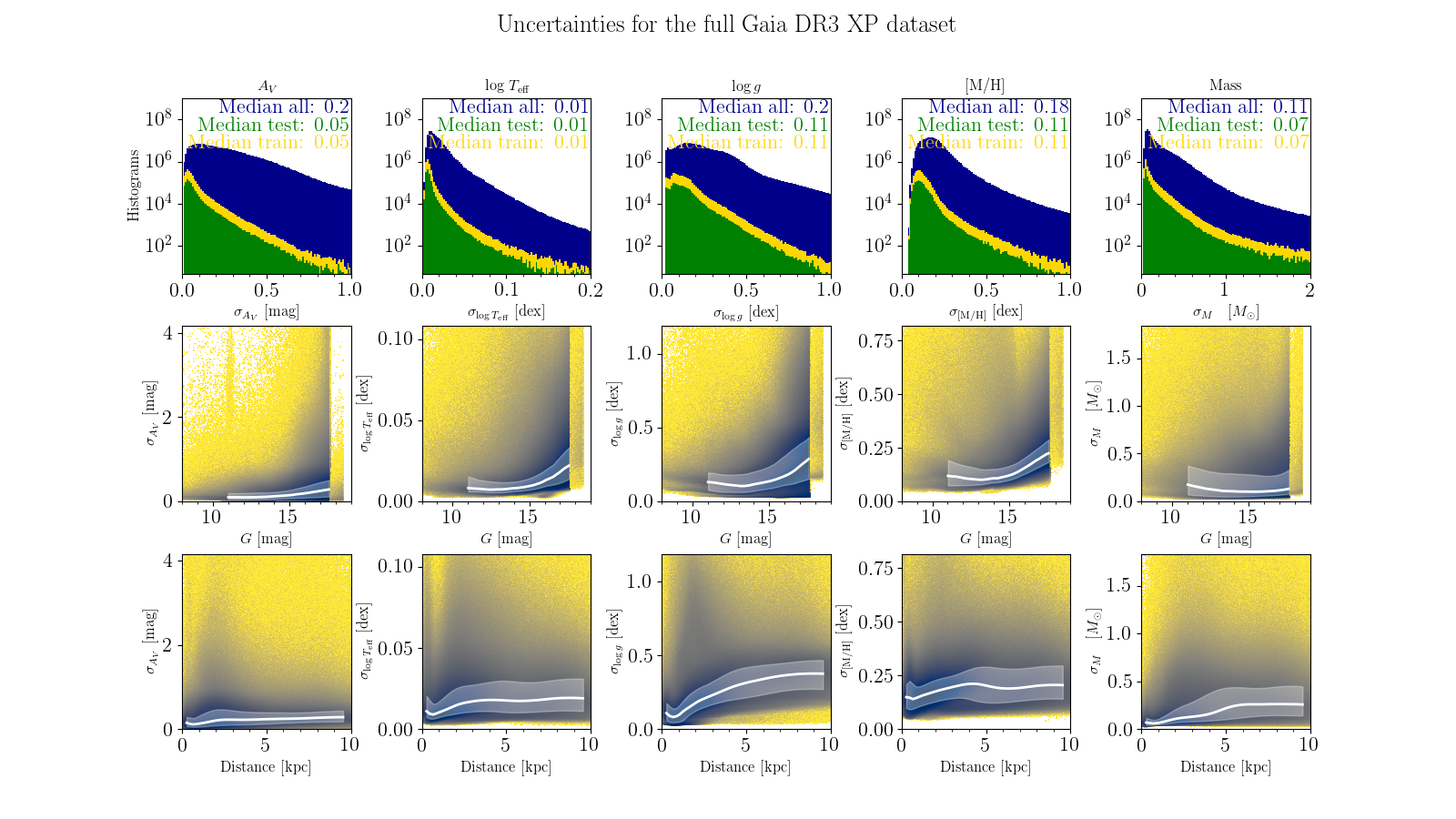}
        \caption{Uncertainty distributions. Top row: Logarithmic histograms of the uncertainties for the training (yellow), test (green), and full {\it Gaia} DR3 XP (blue) samples. The median uncertainties are indicated in each panel. Second row: Uncertainties as a function of {\it Gaia} $G$ magnitude. Bottom row: Uncertainties as a function of distance. In the second and third row, the white lines and shaded bands show the median trends and corresponding 1$\sigma$ quantiles, respectively.}
        \label{fig:validation-hists}
\end{figure*}

\begin{figure*}\centering
        \includegraphics[width=0.33\textwidth]{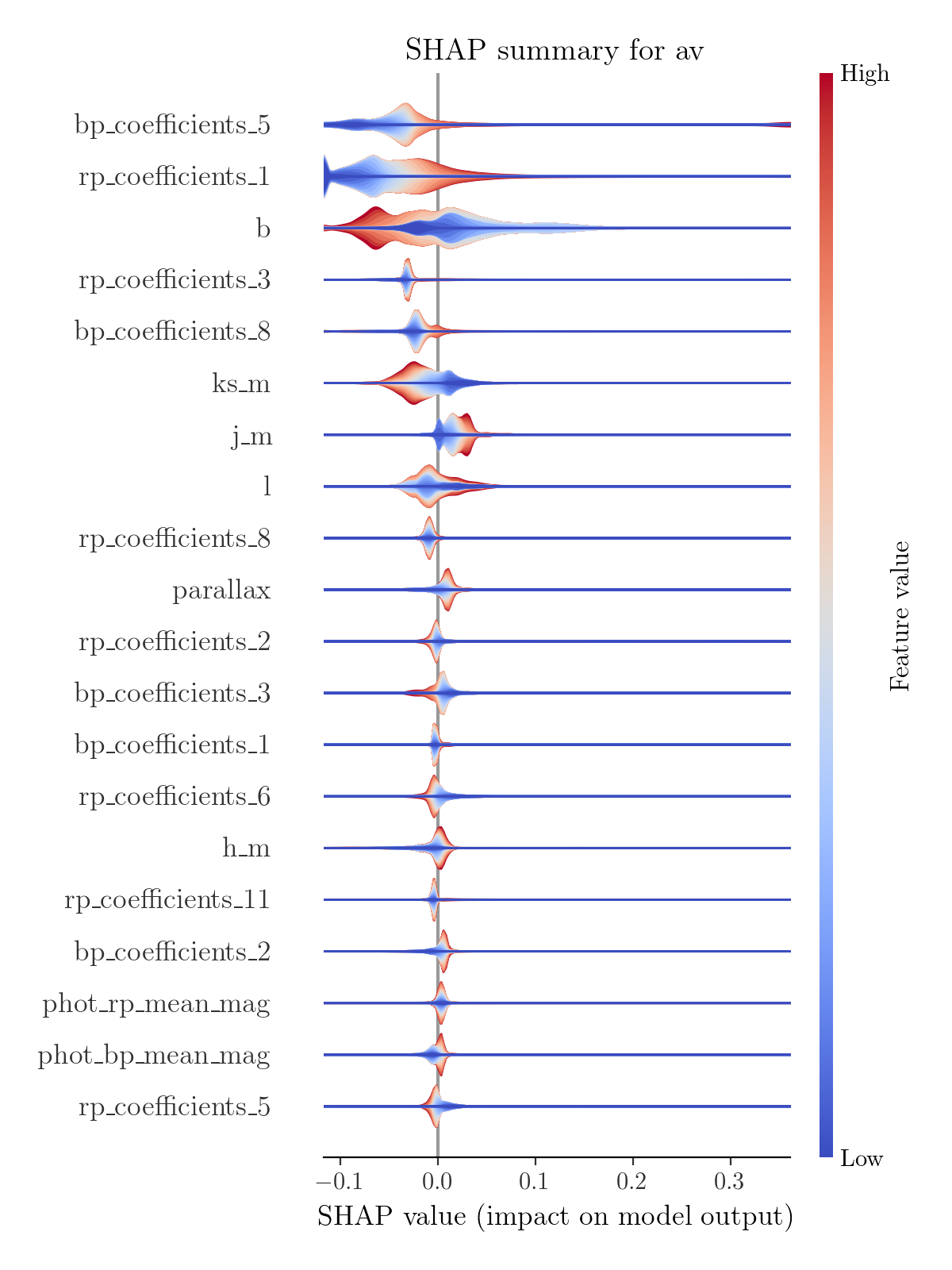}
        \includegraphics[width=0.33\textwidth]{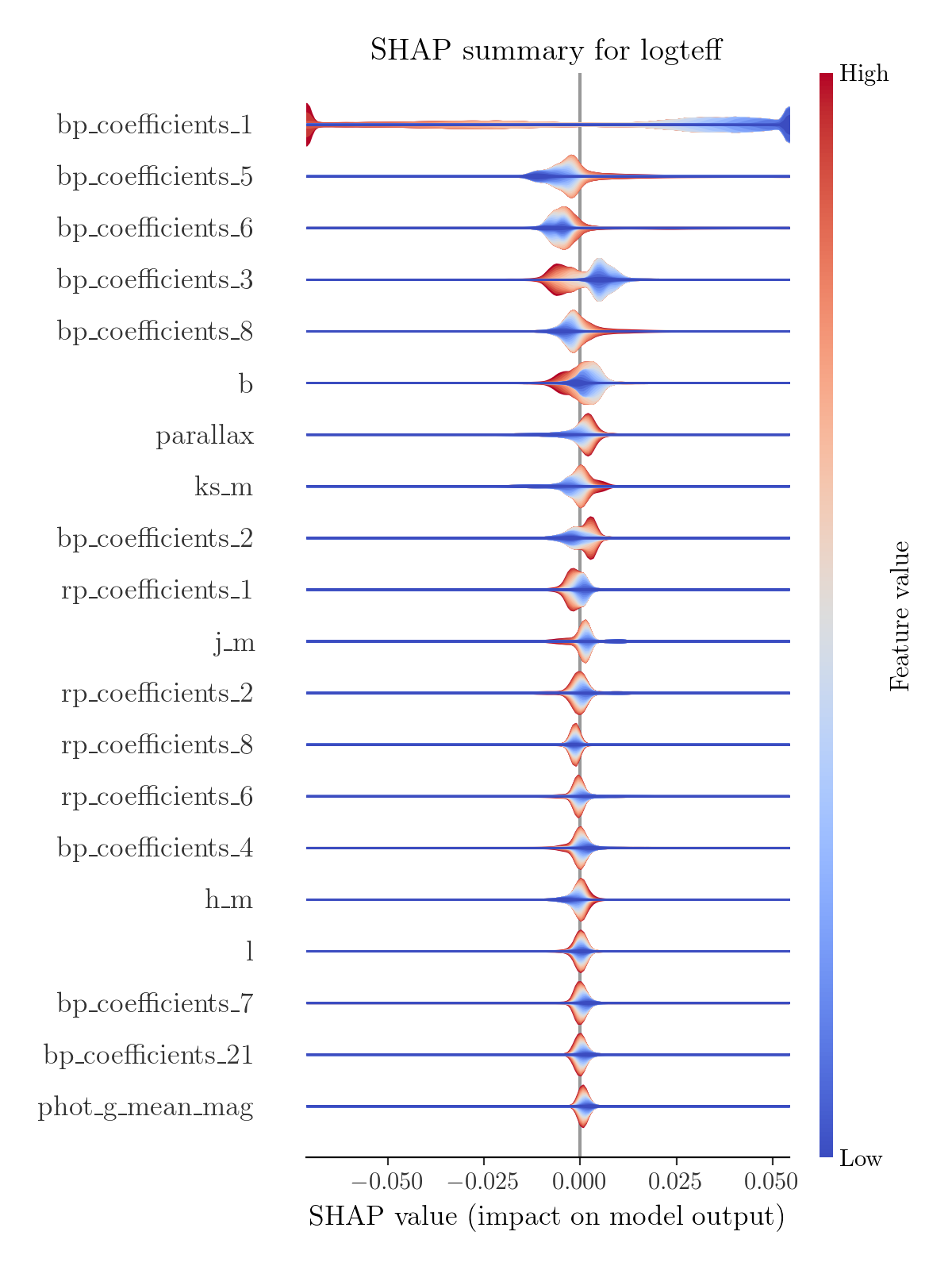}
        \includegraphics[width=0.33\textwidth]{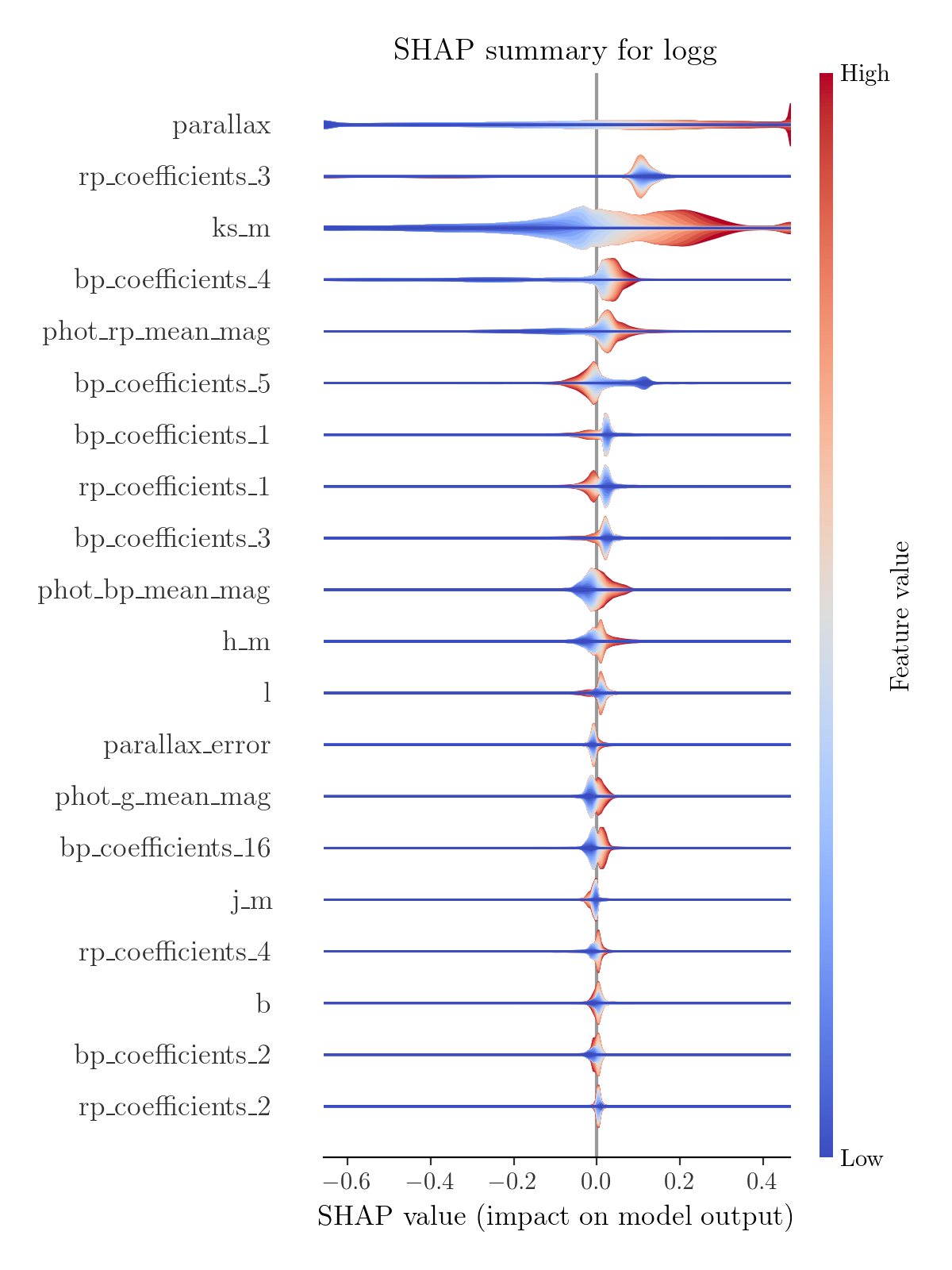}
        \includegraphics[width=0.33\textwidth]{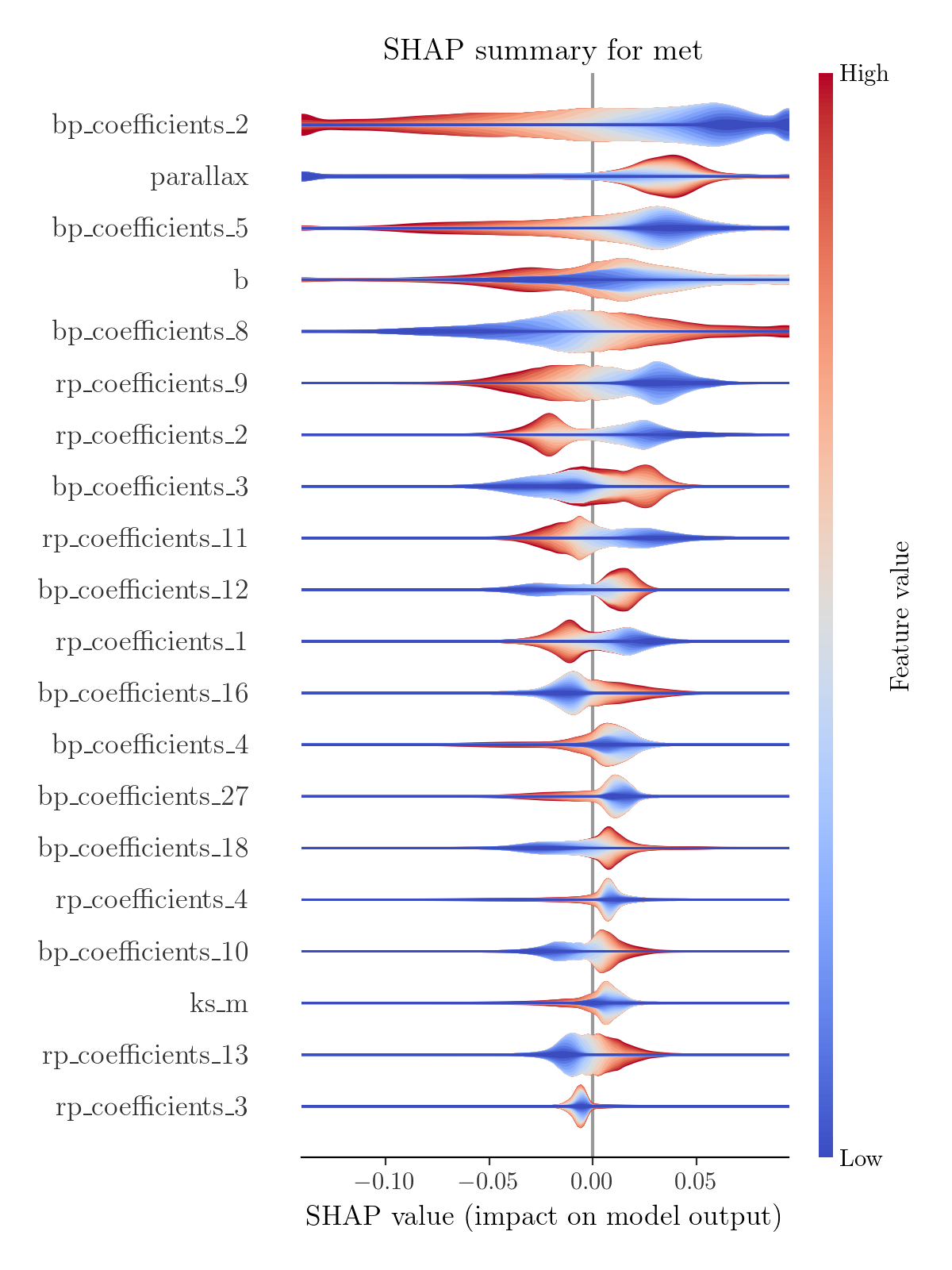}
        \includegraphics[width=0.33\textwidth]{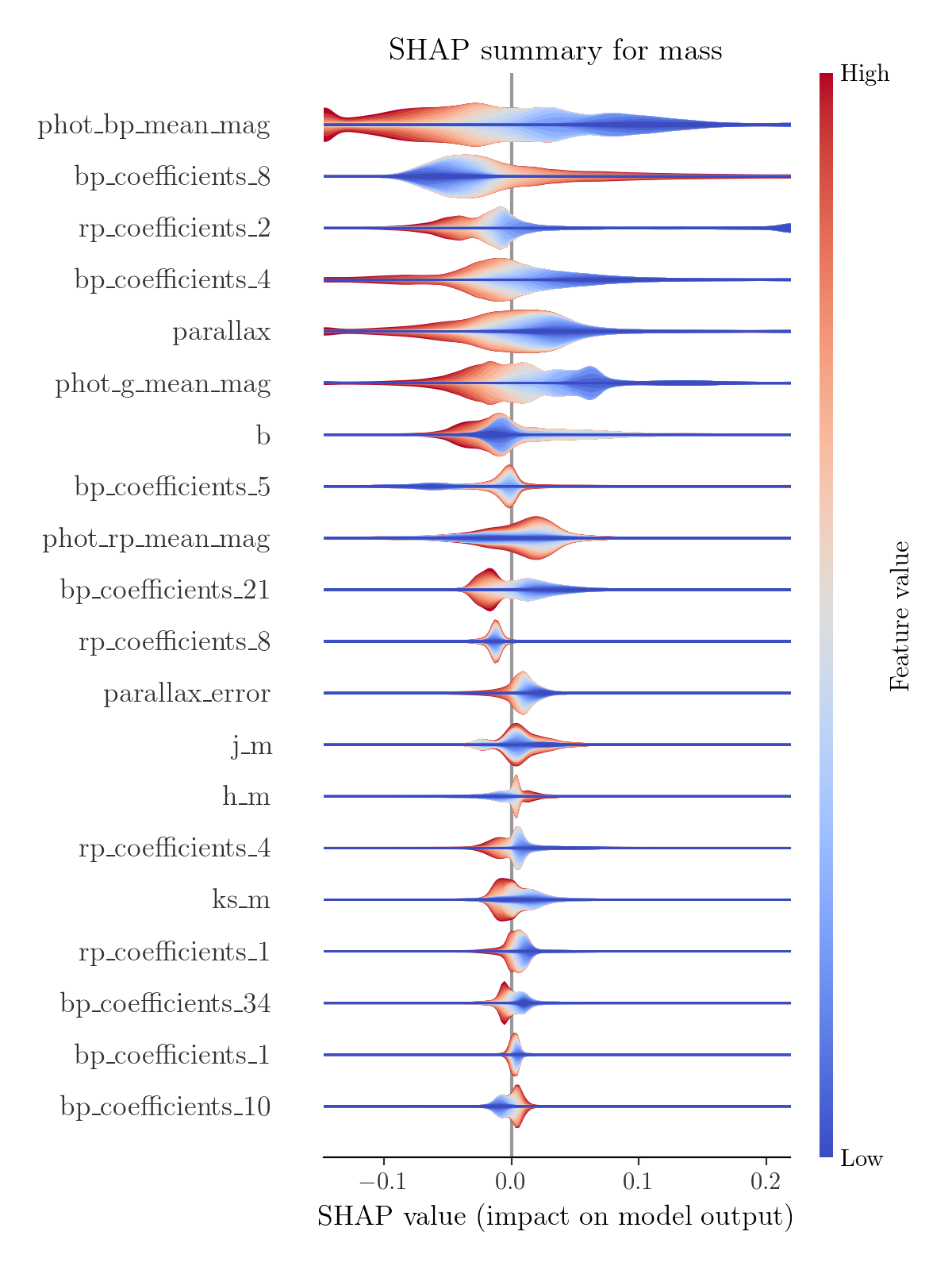}
        \caption{Feature importance plots for each output parameter determined with SHapley Additive exPlanation values (SHAP; \citealt{Lundberg2017}). In each panel, the SHAP values for 20\,000 random stars are aggregated. The rows in each panel are ordered by decreasing feature importance (from top to bottom). Only the 20 most important columns are shown; the abscissa ranges are cut to 99.8\% of the SHAP value range for better visibility.}
        \label{fig:shap}
\end{figure*}

Figure \ref{fig:validation-onetoone} shows the performance of {\tt xgboost} regression on the {\it Gaia} DR3 XP spectra dataset for each of the desirable parameter labels for the test dataset not used in the {\tt xgboost} training phase. The top panels of Fig. \ref{fig:validation-onetoone} show direct comparisons between the labels and the label predictions, the middle panels show the residuals, and the bottom panels the estimated uncertainties. 
Focusing on the middle row, we find that the residuals are typically small in the regions where the bulk of the training and test data are located; furthermore, they typically grow in regimes where training data is sparse (e.g. $\log T_{\rm eff}>3.9$, [M/H]$<-0.8$, $M>1.8{\rm M}_{\odot}$). This behaviour is expected and typical for machine-learning regression techniques, both tree-based and neural networks. It can, in some cases, be mitigated by carefully weighting or balancing the training data \citep[e.g.][]{Ciuca2021} or by defining a different cost function in the fitting process \citep[][Chapters 8.3 and 8.9]{Ivezic2014}. In our case, we tried both attempts but then found worse performances for the full dataset, which in the end led us to keep the unbalanced, but large training set.

The test data uncertainties derived with {\tt xgboost-distribution} are shown in the bottom row of Fig. \ref{fig:validation-onetoone}. They nicely follow the expected trends with the corresponding stellar parameter (e.g. greater $A_V$, $T_{\rm eff}$, or mass values are associated with a greater uncertainty). In particular, the derived uncertainties are always larger than (or at least compatible with) the residuals shown in the middle row of the figure, which shows that the uncertainties are not underestimated, at least for the test dataset.

Some particular features in Fig. \ref{fig:validation-onetoone} may be noteworthy. For example, it appears that the extinction estimates in the low-extinction regime are typically overestimated, but part of this trend stems from (nonphysical) negative extinction estimates in the training data that {\tt xgboost} is compensating for. Another expected feature is the rectangular structure in the $\log g$ comparison (top middle panel in Fig. \ref{fig:validation-onetoone}), which hints at an occasional confusion between dwarfs and red-clump stars in cases where the surface gravity is poorly constrained by the {\it Gaia} parallax. Finally, the thin vertical stripes in the metallicity panels (fourth column in Fig. \ref{fig:validation-onetoone}) are inherited from the training set; namely, the {\tt StarHorse} runs of \citet{Queiroz2023}. A similar (although less pronounced) effect is observed also in the panels that decribe the mass  (last column).

The typical precisions achieved for each label in the full {\it Gaia} DR3 XP sample are presented in Fig. \ref{fig:validation-hists}. In particular, we achieve median precisions of 0.20 mag in $A_V$, 0.014 dex in $\log T_{\rm eff}$ (or 174 K in $T_{\rm eff}$), 0.20 mag in $\log g$, 0.18 dex in [M/H], and 12\% in mass. These uncertainties are highly heteroscedastic: for example, more than 50 million stars (using the 25th percentile of the label uncertainty distributions) have uncertainties smaller than 0.11 mag  in $A_V$, 0.010 dex in $\log T_{\rm eff}$, 0.13 dex in $\log g$, 0.14 dex in [M/H], or 8\% in mass, respectively. As expected, the uncertainty distributions for the full dataset (top row of Fig. \ref{fig:validation-hists}) are broader than for the training and test data (since the bulk of the sample is fainter than most of the training set), but the typical uncertainties are still smaller than the ones obtained from pure astro-photometric isochrone fitting (as performed in e.g. \citealt{Anders2022}; see Appendix \ref{sec:sh2021}). As expected, the reported uncertainties increase with magnitude for most of the labels (except for stellar mass, which is also expected). The uncertainties also increase with distance, as expected.

\subsection{Feature importance}\label{sec:shap}

The concept of SHAP (SHapley Additive exPlanations) values provides an elegant way to understand the output of a machine-learning model \citep{Lundberg2017}. They provide an empirical explanation of how each feature of a given data point impacts the predictions of the model (as also done in \citealt{Anders2023} for the case of spectroscopic stellar age determination). Positive SHAP values mean that the respective feature increases the output’s model value, while a negative SHAP decreases it. The exact behaviour depends on the combination of all other feature values, so that in many cases similar feature values for two particular stars may have opposite effects on the label prediction. For illustrative examples and a detailed explanation, we refer to the documentation of the {\tt shap} package\footnote{\url{https://shap.readthedocs.io/}}.

Figure \ref{fig:shap} shows the SHAP {\tt beeswarm} summary plots\footnote{\url{https://shap.readthedocs.io/en/latest/example_notebooks/api_examples/plots/beeswarm.html}} for each of the five predicted labels, calculated for a random subset of 20\,000 stars. We can think of each star as a pixel in each row. Each subplot shows, from top to bottom, the 20 most influential features (and the exact effect that each of them has) in the prediction of the corresponding label ($A_V, T_{\rm eff}, \log g,$ [M/H], and $M$). It is clear that the {\it Gaia} XP coefficients contain very valuable information for the prediction of the stellar labels. For example, for $A_V$ the most important features are the fifth BP coefficient and the first RP coefficient, followed by Galactic latitude (which is obviously a good first prior for the presence of dust), the third RP coefficient, the eighth BP coefficient, and the 2MASS $K_s$ and $J$ magnitudes.
In the case of effective temperature (second panel in Fig. \ref{fig:shap}), we observe that the first BP coefficient (obviating the zeroth coefficient that was used to normalise the XP spectra) is the most influential piece of information (in accordance with e.g. Figs. 25 and 26 in \citealt{DeAngeli2023}). For most labels,  the {\it Gaia} parallax also plays an important role in the determination of the output.

Recently, \citet{Guiglion2024} also explored feature importance using neural network gradients in the context of their hybrid-CNN method for the {\it Gaia} RVS spectra. The hybrid-CNN method employs an entirely different machine learning approach and input dataset, with a combination of {\it Gaia} DR3 RVS spectra, photometry ($G, G\_{BP}, G\_{RP}$), parallaxes, and XP coefficients. They demonstrated that for each label (i.e. $T_{\rm eff}$, $\log g$, [M/H], [$\alpha$/Fe], [Fe/H]) their neural network learns from the characteristic spectral features in the {\it Gaia} RVS spectra and the specific {\it Gaia} XP coefficients, in addition to the parallax and magnitudes. The difference in the machine learning methodology, the dataset and the method of estimation of the feature importance can result in a different relative importance of the features.

For $T_{\rm eff}$ values found by the hybrid-CNN, in decreasing order of importance, the fifth, fourth, second, sixteenth, and first BP coefficients provide the most information. For $\log g$, the fourth, sixteenth, second, fifth and eighth BP coefficients are found to be important.  For [M/H], the sixteenth BP coefficient was found to be most important followed by the  fourth, sixth, fourteenth and eleventh RP
coefficients. The parallax information is also found to be very crucial for $\log g$ and $T_{\rm eff}$ (as previously demonstrated by \citealt{Guiglion2020} for the RAVE spectra). The most important XP coefficients identified for the hybrid-CNN method (as listed above and also see their Fig. 10) are found to be common with most important XP coefficients we present in Fig. \ref{fig:shap} for the {\tt xgboost} method.
A more detailed comparison with the stellar parameter catalogue provided by \citet{Guiglion2024} is given in Appendix \ref{sec:gui2024}.

\section{The {\tt SHBoost} {\it Gaia} DR3 catalogue}\label{sec:catalogue}

In this paper, we  use an {\tt xgboost} regression to produce a catalogue of stellar properties derived from {\it Gaia} DR3 XP spectra, astrometry, and multi-wavelength photometry. This catalogue, referred to as {\tt SHBoost}, comprises the extinction, effective temperature, surface gravity, [M/H], and mass estimates for more than 217 million stars and is made available via CDS/VizieR. The data model is described in Appendix \ref{sec:datamodel}.

As mentioned in Sect. \ref{sec:errors}, Fig. \ref{fig:validation-hists} shows a summary plot of the uncertainties for each parameter, as histograms (top row) and as a function of magnitude (second row) and distance (bottom row), respectively. Below, we give a few examples of how our catalogue can be used to study stellar populations and interstellar dust in the Milky Way and the Local Group. Comparisons with other stellar-parameter catalogues based on {\it Gaia} DR3 are presented in Appendix \ref{sec:literature}. 

\subsection{Kiel and colour-magnitude diagrams}\label{sec:cmds}

\begin{figure}
        \includegraphics[width=0.49\textwidth]{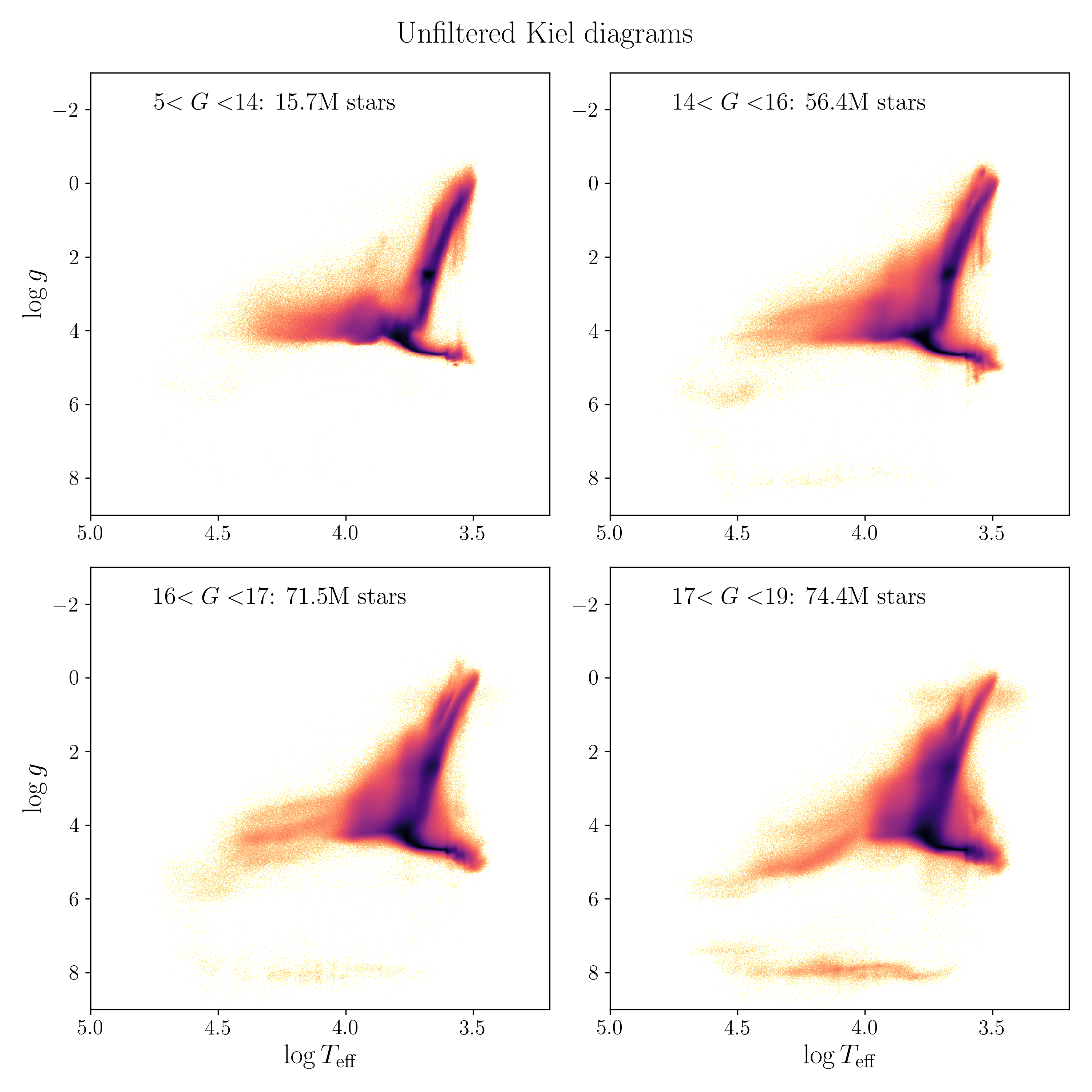}
        \caption{Unfiltered {\it Kiel} diagrams of the full {\it Gaia} DR3 XP sample, in four broad bins of observed $G$ magnitude.}
        \label{fig:kiel}
\end{figure}

\begin{figure*}\centering
        \includegraphics[width=0.9\textwidth]{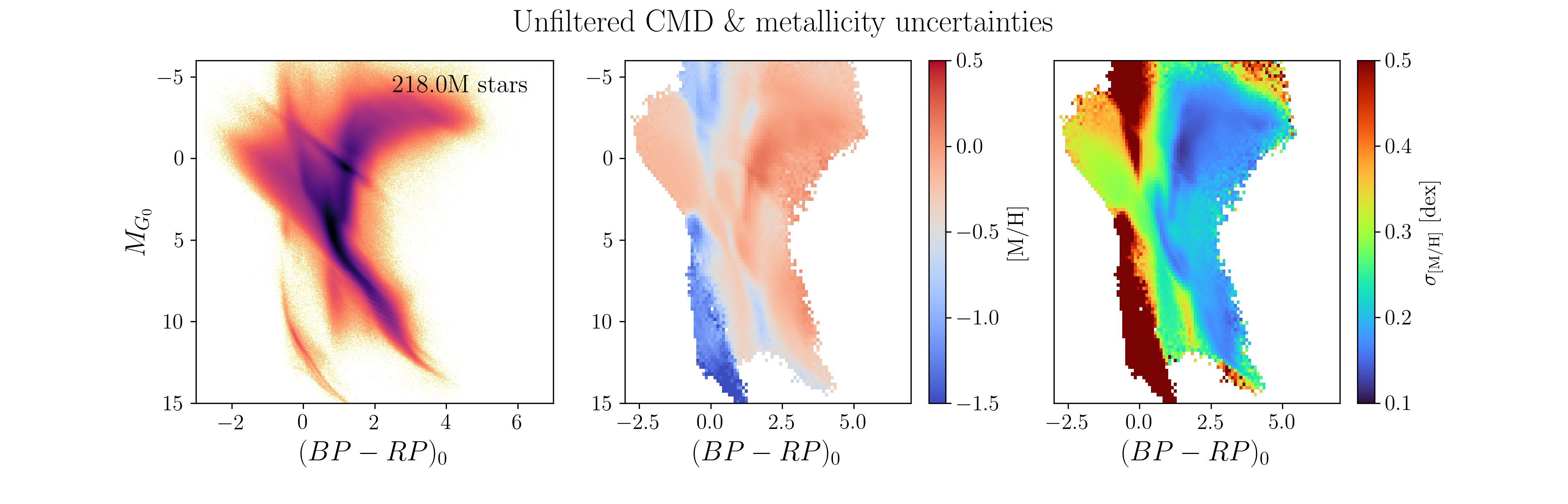}
        \includegraphics[width=0.9\textwidth]{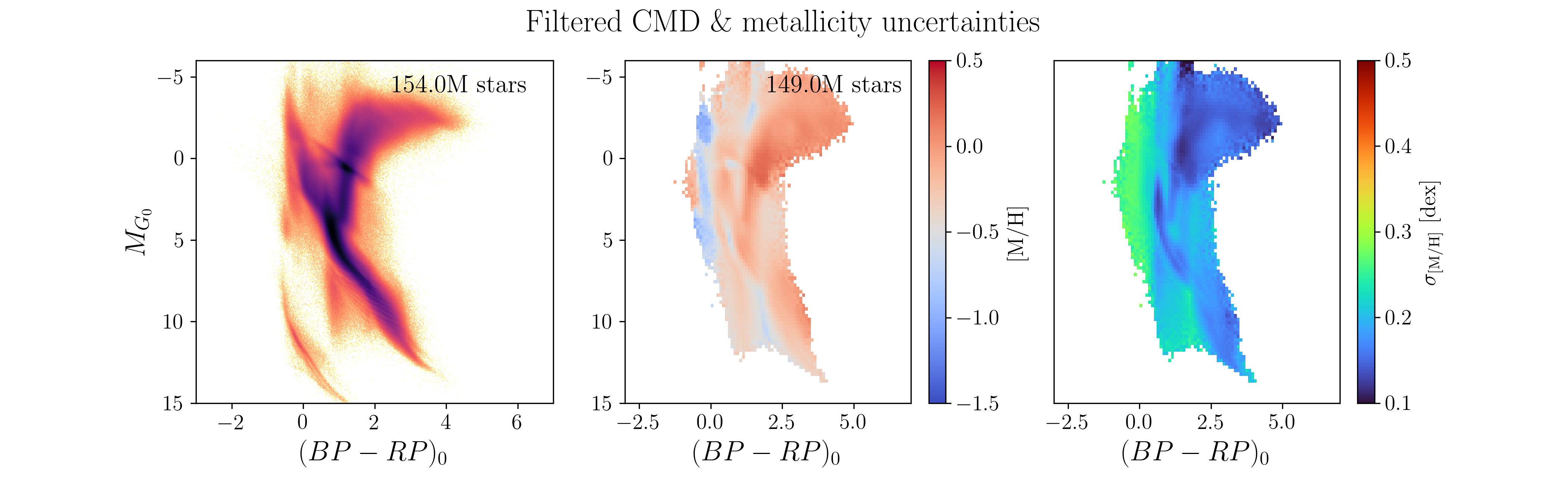}
        \includegraphics[width=0.9\textwidth]{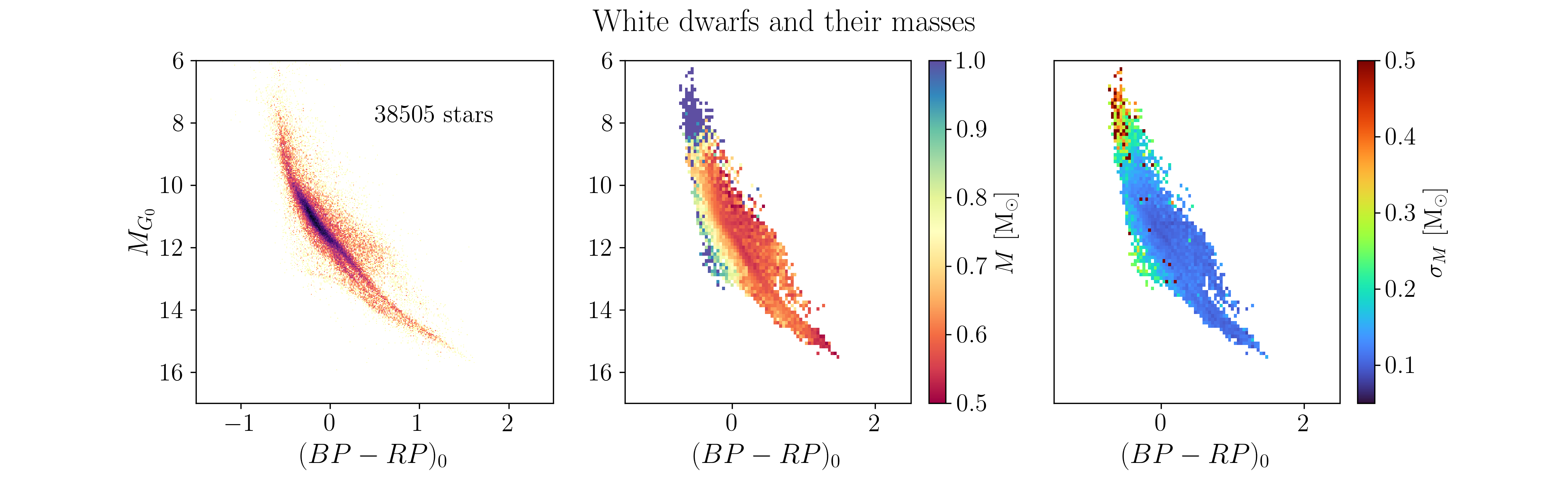}
        \caption{Extinction-corrected colour-magnitude diagrams. Top row: Full {\it Gaia} DR3 XP sample (before filtering results with large uncertainties). Top left: 2D histogram. Top middle: colour-coded by median metallicity per bin. Top right: Colour-coded by median metallicity uncertainty per bin. Middle row: Same as top row, but flag-cleaned (using {\tt xgb\_av\_outputflag}, {\tt xgb\_logteff\_outputflag}, and {\tt xgb\_met\_outputflag}). Bottom row: White-dwarf portion of the colour-magnitude diagram (filtered by $\log g > 7$). Bottom left: 2D histogram. Bottom middle: Mass colour-coded by median mass. Bottom right: Mass colour-coded by mass uncertainty.}
        \label{fig:cmds}
\end{figure*}

In Fig.~\ref{fig:kiel} ,we show the {\it Kiel} diagrams ($\log g$ vs $\log T_{\rm eff}$) obtained from the full {\it Gaia} DR3 XP run in four magnitude bins ($G < 14$, $14 < G \leq 16$, $16 < G \leq 17$, and $17 < G \leq 19$). Figure \ref{fig:cmds} shows the colour-magnitude diagrams (corrected for distance and extinction) for the full {\it Gaia} DR3 XP sample for which $G_{BP}$, $G_{RP}$, and parallaxes are available (217.97M stars) and for a sub-sample of white-dwarf stars (one of the novelties of the present work with respect to \citealt{Anders2022}).

At a first glance, the diagrams are unsurprising and show expected trends (for example, the {\it Kiel} diagrams in the four magnitude bins shown in Fig. \ref{fig:kiel} look similar, and the full CMD in Fig. \ref{fig:cmds} resembles the one in the lower right panel of Fig. \ref{fig:trainingset_cmds}). However, upon a closer inspection, we notice a series of interesting details.

The top panels of Fig. \ref{fig:kiel} show {\it Kiel} diagrams that are almost fully consistent with our expectations from stellar evolutionary models. As our composite training set covers most of the areas of {\it Kiel} diagram actually occupied by the Milky Way's stellar populations and is indirectly based on stellar models, this is expected. However, since our predictions of $T_{\rm eff}$ and $\log g$ are independent (in the sense that we train an {\tt xgboost} model for each label separately), this is not a completely trivial result. In fact, the upper right panel, for example, shows that some stars around $\log T_{\rm eff}\simeq3.6$ fall in unphysical $\log g$ ranges, which indicates that our capacity to determine their surface gravity is not at the same level as our $T_{\rm eff}$ accuracy. 

The brightest magnitude bin (upper left panel of Fig. \ref{fig:kiel} does not contain any WDs, since these are too intrinsically faint. The bulk of the WDs is contained in the faintest magnitude bin (beyond the magnitude limit of $G=17.65$ for non-WDs in {\it Gaia} DR3; \citealt{Montegriffo2023}). Nevertheless, the total number of white dwarfs contained in the XP dataset is low (see also Fig. \ref{fig:cmds}; lower row).

The second magnitude bin (upper right panel of Fig. \ref{fig:kiel}) contains a small number of stars classified as white dwarfs ($\log g \simeq 8$) and hot sub-dwarfs ($\log g \simeq 6$), the latter well separated from hot main-sequence stars ($\log g \simeq 4$). This separation becomes less obvious in the fainter magnitude bins (the hot part of the MS tilts towards higher $\log g$ in the lower panels of Fig. \ref{fig:kiel}), which means that our results do not allow for a completely clean separation between O-type main-sequence stars and hot sub-dwarfs (consistent with the findings of \citealt{GaiaCollaboration2018B}).

The lowest-mass MS is missing in our training set, which results in overestimated effective temperatures for ultra-cool dwarfs. These objects accumulate at the lower end of the temperature sequence in Fig. \ref{fig:kiel}.
 
The top left panel of Fig. \ref{fig:cmds} shows that there is a sizeable number of (mainly red-clump, but also upper-MS) stars with inaccurate/uncertain extinction estimates, which results in the unphysical diagonal stripes (the most obvious one extending above and below the red clump). This feature is also visible in parts of the training set; in particular in the {\tt StarHorse} LAMOST LRS and MRS data (central panels of Fig. \ref{fig:trainingset_cmds}). We tested the exclusion of these data from the training set, but achieved worse results -- indicating that the fainter magnitude limit of LAMOST makes these stars crucial during the training phase. We thus look forward to a re-calibration of all spectroscopic survey data to a common stellar parameter scale (see e.g. the recent efforts of \citealt{Thomas2024}), which should lead to a significant improvement of the stellar-parameter homogenisation problem in the future.

The second and third panels of Fig. \ref{fig:cmds} show the intrinsic CMD colour-coded by metallicity (median {\tt xgbdist\_met\_mean} per pixel) and metallicity uncertainty (median {\tt xgbdist\_met\_std} per pixel), respectively. In particular, the right panel highlights the CMD regions in which the XP spectra are most sensitive to metallicity (FGK dwarfs and giants, as expected). The derived metallicities of WDs and stars in edge regions of the {\it Kiel} diagram/CMD should not be used.

\subsection{Caveats}\label{sec:caveats}

In this subsection we reiterate the observational and methodological caveats associated with our {\tt SHBoost} catalogue. Some of them were already discussed with the validation plots shown in the previous subsections or in the comparisons to other stellar-parameter catalogues shown in Appendix \ref{sec:literature}. The most important caveats of our method are described in this section.

Most importantly, the accuracy of any supervised machine learning model is limited by the accuracy of the training set it uses. In our case, most of our training set is coming from the {\tt StarHorse} catalogues of \citet{Queiroz2023} derived from ground-based spectroscopic surveys and {\it Gaia} DR3. These stellar parameters are based on state-of-the-art evolutionary models, but do not take into account unresolved or interacting binaries.

A second (and related) caveat is that in order to cover a larger parameter space of the Hertzsprung-Russell diagram, we joined the \citet{Queiroz2023} catalogues with other catalogues of hot stars, white dwarfs, very metal-poor stars, and hot sub-dwarfs (see Fig. \ref{fig:trainingset_cmds}). At the margin between these regimes, systematics can be expected. These systematics certainly contribute to both the statistical and the systematic uncertainty budget of the {\tt SHBoost} catalogue and, thus, the dedicated catalogues for smaller parts of the Hertzsprung-Russell diagram will likely be more precise within their specified realm of validity.

A minor caveat in this respect is that there are also systematic differences in the input spectroscopic stellar parameters of the different {\tt StarHorse} catalogues. Our training set is based on the {\tt StarHorse} posteriors, so the systematics between different surveys are to a large degree homogenised, but some systematics might still be present (see discussion in \citealt{Queiroz2023}). A true inter-survey homogenisation should follow approaches such as \citet{Ness2015, Ting2019, Guiglion2020, Ambrosch2023}, or \citet{Thomas2024}. Homogeneous stellar parameter catalogues might become available for more surveys in the near- and mid-term future \citep{Tsantaki2022}.

Due to our philosophy of using the {\tt StarHorse} posteriors as the input label scale, the produced {\tt xgb\_met} metallicities are based on [M/H] (assuming the Salaris correction for $\alpha$-enhancement). Thus, we see significant trends with respect to [Fe/H] values, particularly at low metallicities (see Appendix \ref{sec:feh_calib}). Although we included a significant set of very metal-poor stars in the training data, our catalogue results also comes with much larger uncertainties at low metallicities.

By construction, {\tt xgboost}, unlike {\tt StarHorse} or similar Bayesian stellar inference codes, cannot take into account co-variances between the stellar parameters. Correlations between, for example, $T_{\rm eff}$ and $\log g$ are therefore not taken into account. This weakness cannot be overcome with currently available tree-based methods, while it is intrinsically taken care of by many multi-label output neural networks (e.g. \citealt{Fallows2022, Fallows2024, Anders2023Proc, Guiglion2024}). Nevertheless, we have achieved physically meaningful results for the {\it Kiel} diagram (see Fig. \ref{fig:kiel}), for instance. This means that many stellar-parameter degeneracies that appear in purely photo-astrometric inferences (e.g. \citealt{Anders2019, Anders2022}) disappear when using the {\it Gaia} XP spectra. 

We obtained stellar parameter and extinction estimates for all objects with {\it Gaia} DR3 XP spectra (217\,974\,770). This sample includes also a small portion of bright ($G<17.65$) galaxies and quasars. For a significant portion of stars, some of the derived quantities are very uncertain or even meaningless (see e.g. the metallicities of white dwarfs in Fig. \ref{fig:cmds}). All these results should be filtered out either by their uncertainties or the corresponding parameter flags.

In this work, we did not attempt to derive new age or distance estimates, since the information about these quantities is not directly encoded in the {\it Gaia} DR3 XP spectra. All the maps in this paper use Bayesian distance estimates from \citet{Anders2022} or \citet{Bailer-Jones2021}, both of which are based on {\it Gaia} (E)DR3 astrometry and photometry.

\section{Results}\label{sec:results}

\subsection{Metallicity maps}\label{sec:maps}

\begin{figure*}\centering
        \includegraphics[width=0.49\textwidth]{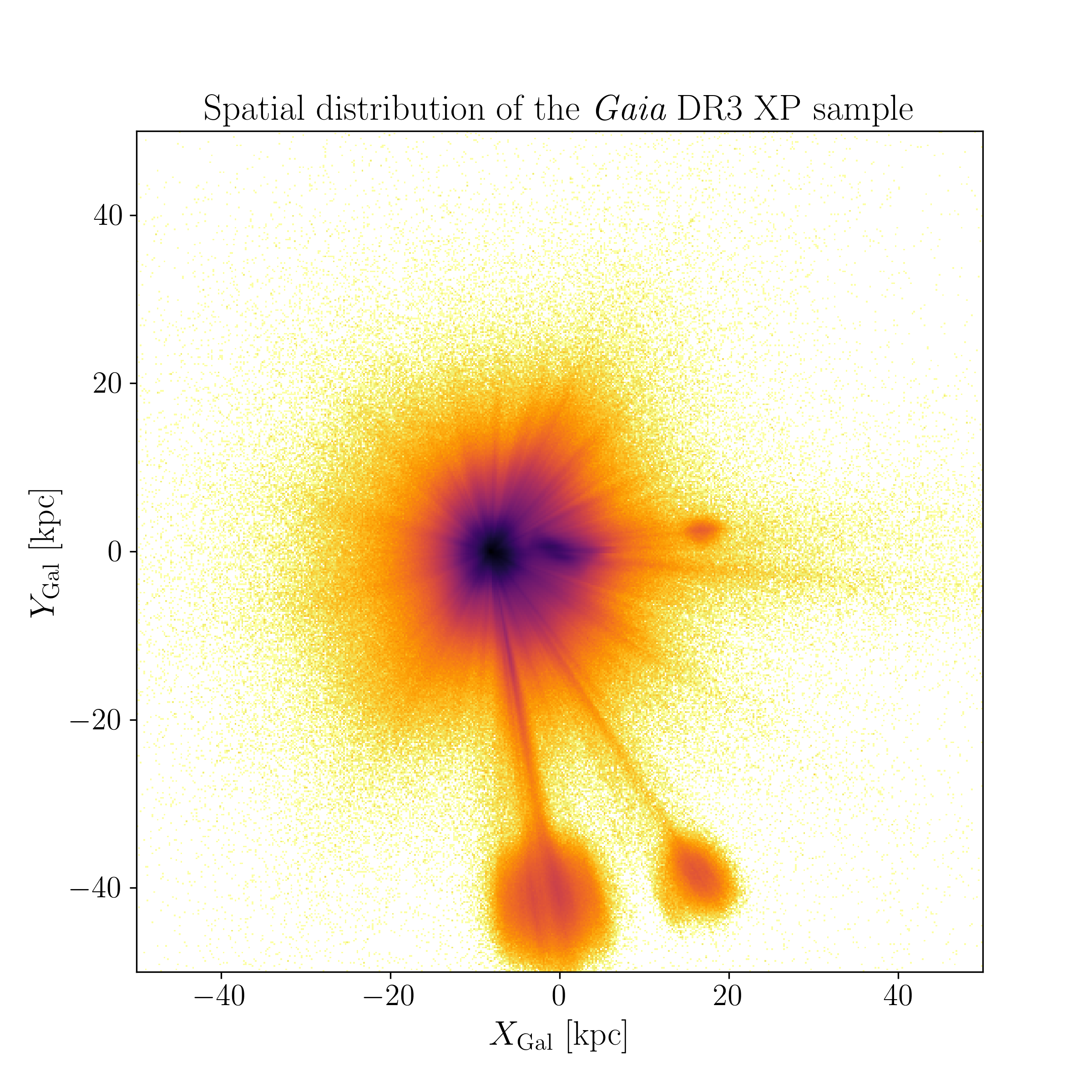}
        \includegraphics[width=0.49\textwidth]{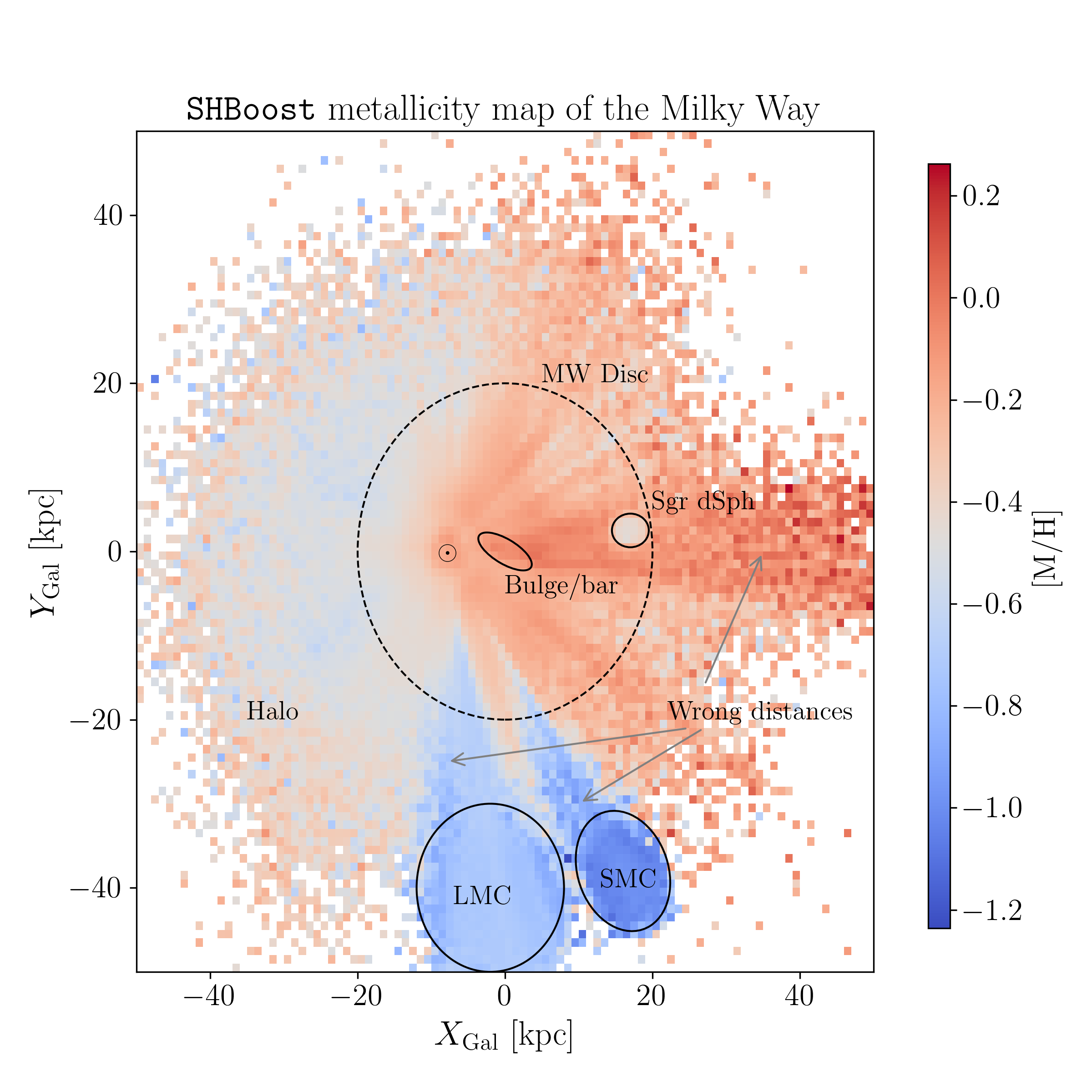}
        \caption{Large-scale Cartesian map (box size: 100x100 kpc$^2$) of the {\it Gaia} DR3 XP sample. Left panel: Density map. Right panel: Median metallicity map of the same volume (integrating over all $Z_{\rm Gal}$), using all 204 million stars with $\sigma_{\rm [M/H]}<0.3$ dex and only showing bins containing more than 3 stars. Some salient features of the map are annotated.}
        \label{fig:metallicity_map}
\end{figure*}

\begin{figure}\centering
        \includegraphics[width=0.49\textwidth]{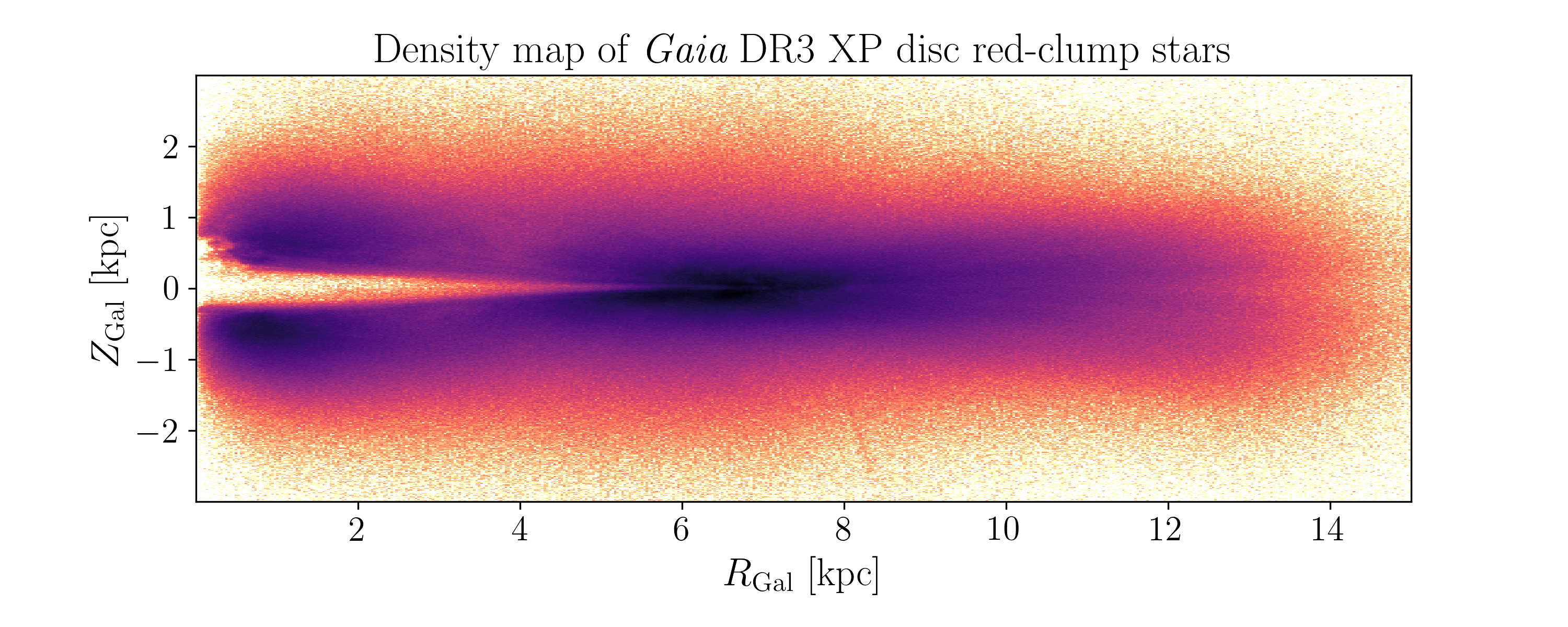}
        \includegraphics[width=0.49\textwidth]{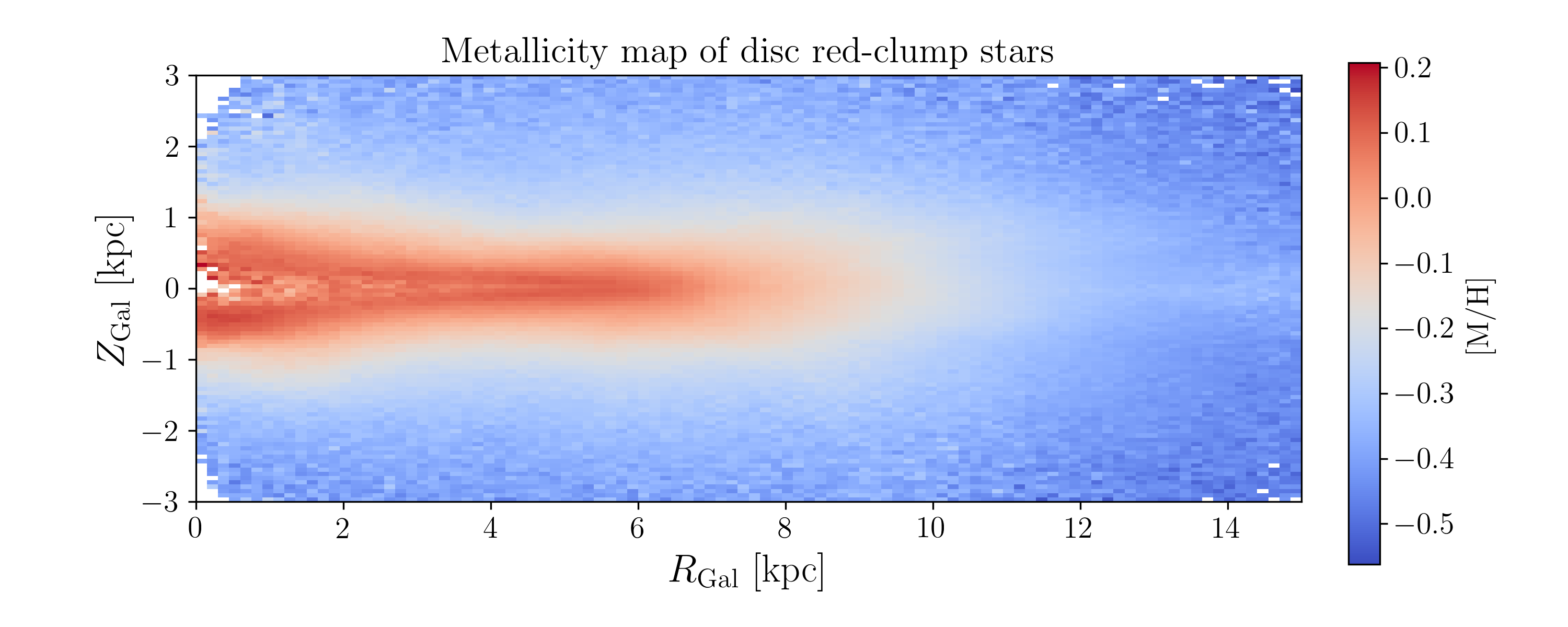}
        \caption{Distribution of red-clump stars with {\it Gaia} DR3 XP spectra in the Galactic disc. Top panel: Density distribution in cylindrical Galactocentric coordinates. Bottom panel: Median metallicity map of 7.5 million red-clump stars with $\sigma_{\rm [M/H]}<0.2$ dex.}
        \label{fig:metallicity_redclumps}
\end{figure}

Figures \ref{fig:metallicity_map} and \ref{fig:metallicity_redclumps} show examples of metallicity maps that can be created from our dataset. The left panel of Fig. \ref{fig:metallicity_map} shows a face-on map of the Milky Way and its surroundings from afar (box size: 100 kpc x 100 kpc), based on distances from \citet{Anders2022} when available, otherwise from \citet{Bailer-Jones2021}. The right panel of the figure shows the same map (but colour coded by the median {\tt xgbdist\_met\_mean} per pixel) demonstrating that it is possible to see some astrophysical signatures in the data (e.g. we appreciate the expected metallicity difference between the Large and Small Magellanic Clouds), while others have not been recovered (e.g. a negative metallicity gradient in the MW halo; e.g. \citealt{Castellani1983, Carney1990, TuncelGuctekin2019}, but see also \citealt{Monachesi2016, Das2016, Conroy2019, Vickers2021}). We suggest that this is partly due to uncertain distance estimates and partly due to the poor performance of the {\tt xgbdist\_met} estimator in the low-metallicity regime, which can be mitigated by using calibrated [Fe/H] estimates (see Sect. \ref{sec:feh_calib}).

In Fig. \ref{fig:metallicity_redclumps}, we show an edge-on view of the Milky Way disc ($R_{\rm Gal}$ vs $Z_{\rm Gal}$), using only red-clump stars with small {\tt xgbdist\_met\_std}. The median metallicity map (bottom panel) displays the expected signatures of the radial and vertical metallicity gradients, covering a much larger range of the disc with good statistics than currently possible with high- or medium-resolution spectroscopic surveys. A detailed quantitative comparison is out of the scope of the present paper, but the plot clearly reaffirms that the information content of the low-resolution {\it Gaia} XP spectra should not be underestimated (see also e.g. \citealt{Andrae2023, Lucey2023, Weiler2023, Li2024}).

\subsection{Hot stars}\label{sec:hot}

\begin{figure}\centering
        \includegraphics[width=0.49\textwidth]{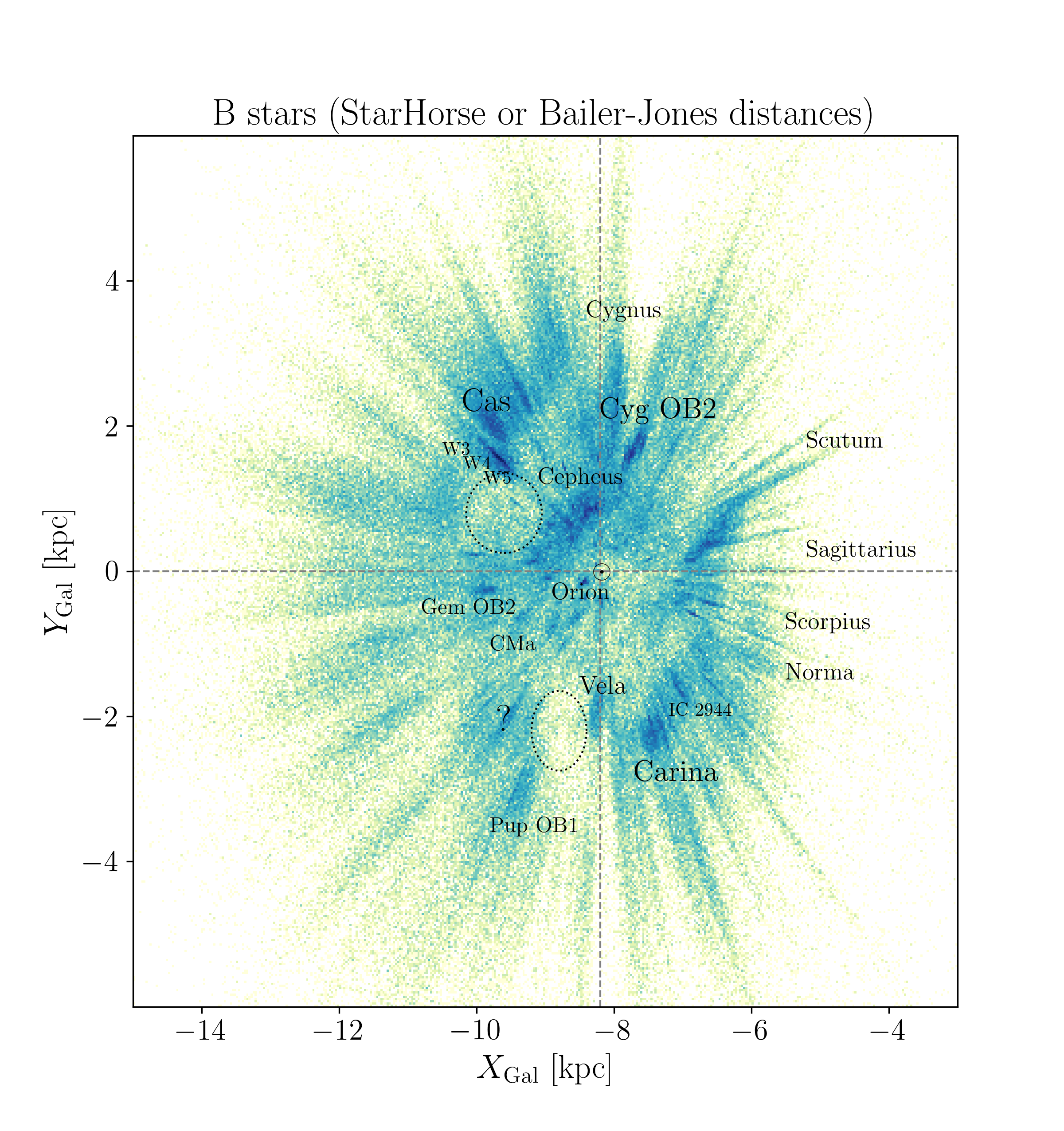}
        \caption{Spatial distribution of a {\it Gaia} DR3 XP sample of $376\,321$ B star candidates (selected as $4.0 <$ {\tt xgbdist\_logteff\_mean} $< 4.5\, \& $ {\tt xgbdist\_logg\_mean} $< 6\, \& ${\tt xgbdist\_logteff\_std} $< 0.1\, \& $ {\tt xgbdist\_logg\_std} $< 0.5$). Known over-densities corresponding to OB associations are annotated. The dotted ellipses correspond to (potential) star formation voids.}
        \label{fig:b_stars}
\end{figure}

One of the strong points of our method lies in the diversity of the training set. Since we explicitly included hot stars in our training set, the stellar-parameter predictions for these stars are significantly better than in \citet{Anders2022} or \citet{Andrae2023G}. As an example, we show in Fig. \ref{fig:b_stars}, a map of about 375\,000 B star candidates selected only based on their effective temperature and surface gravity. This map extends further than most previous maps of hot stars (e.g. \citealt{Zari2021, Pantaleoni2021}) and provides fully consistent results with the {\it Gaia}-derived maps of young upper main-sequence stars \citep[UMS,][]{Poggio2021, GaiaCollaboration2023D}.

Figure \ref{fig:b_stars} demonstrates that even without a dedicated effort to determine better distances for these stars (which would certainly be necessary for further science exploitation), all known major over- and under-densities have been recovered (the plot can be compared to e.g. Fig. 11 in \citealt{Zari2021} or Fig. 5 in \citealt{Pantaleoni2021}). In addition, the new over-densities in the Milky Way's fourth quadrant seen in the {\it Gaia} EDR3 maps of \citet{Poggio2021} (at $(-9.5, -2)$ or at $(-9, -3)$) have been confirmed. As pointed out in \citet{GaiaCollaboration2023D}, they also coincide  with over-densities of young open clusters seen in recent catalogues \citep{Cantat-Gaudin2020, Hunt2023}.

The hot star map shown in Fig. \ref{fig:b_stars} also highlights two under-densities (dashed ellipses) that appear in many different samples of young stellar tracers (star clusters, O stars, upper main-sequence stars; e.g. \citealt{Castro-Ginard2019})\footnote{\url{https://www.cosmos.esa.int/web/gaia/iow_20180614}}. In both cases, we find a significant number of sources that have been detected in the same line of sight but further away; in other words, they have not been produced by extinction cones. Recent 3D extinction maps (e.g. \citealt{Vergely2022}, Fig. 11) also indicate that there is little absorption in the respective areas. The sample of IPHAS-selected A stars investigated by \citet{Ardevol2023} is limited to longitudes $30\deg<l<215\deg$; thus, it does not reach this region.

Comparing Fig. \ref{fig:b_stars} to the map of {\it Gaia} EDR3 upper main-sequence stars (\citealt{Poggio2021}, see Fig. 1), we find that both these two over-densities and the under-density near Vela can also be sensed, albeit close to the limit of their sample. This is a further point indication that these features are not biases of this new sample. We argue that the mapping of stellar under-densities should be just as important as mapping their overdensities. We see more and more evidence for lumpy or flocculent structure in the solar neighbourhood, as opposed to a smooth distribution with well-defined spiral arms \citep[e.g.][]{Shetty2008, Dobbs2011, Dobbs2014, Castro-Ginard2021}.

\subsection{Extinction maps}\label{sec:extinction}

\begin{figure}\centering
        \includegraphics[width=0.49\textwidth]{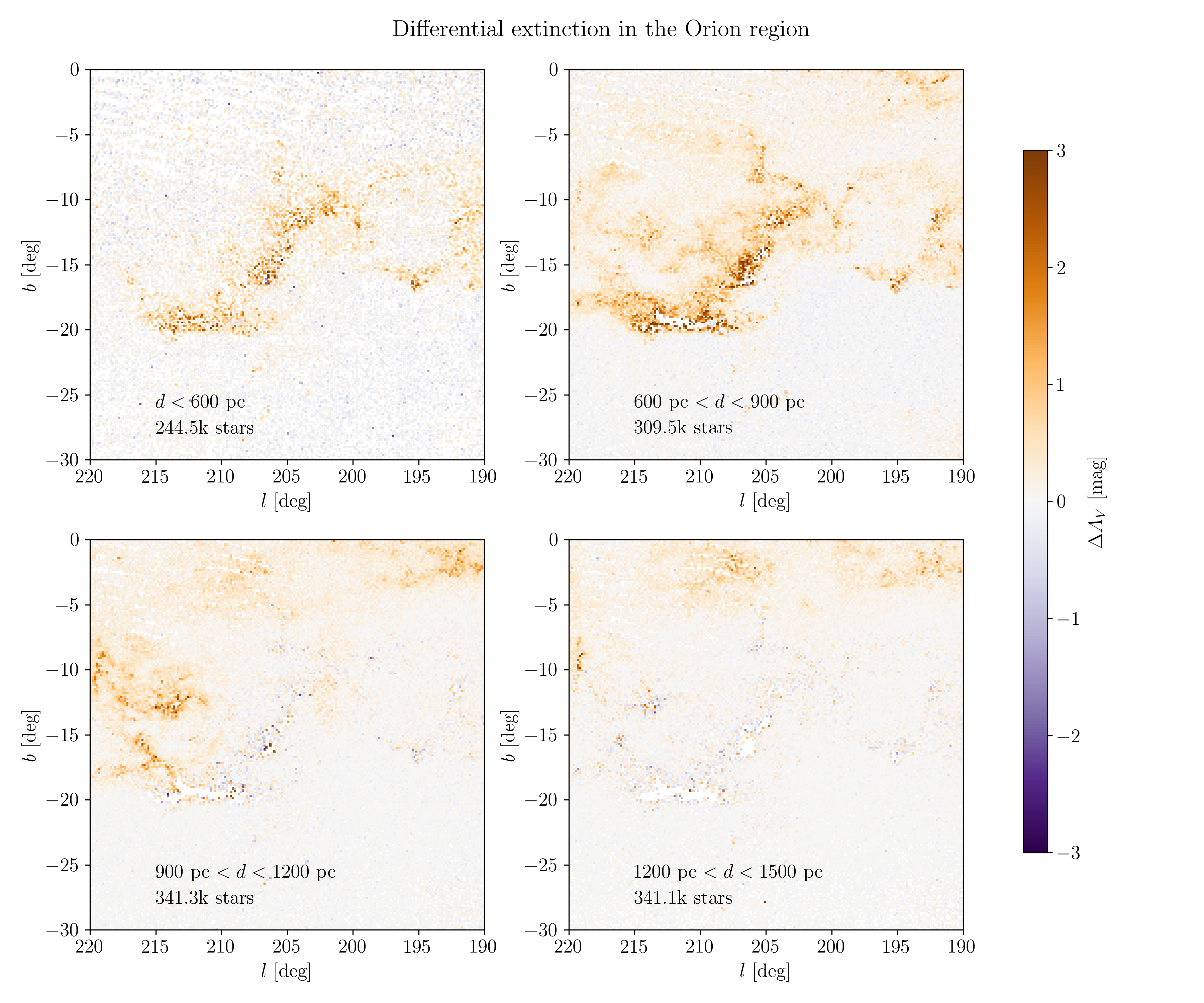}
        \caption{Differential $\Delta A_V$ extinction maps for the Orion region (30 deg x 30 deg, $d<1.5$ kpc; left) in four bins of distance, as indicated in each panel. Some artefacts from the {\it Gaia} scanning law are visible in these maps, but  the distance-sliced $A_V$ maps are generally complete down to $A_V\gtrsim6$ mag, with a high resolution, and correspond very well to existing 3D dust maps.}
        \label{fig:orion}
\end{figure}

Figure \ref{fig:orion} shows an example of how our results perform in terms of extinction. The four panels correspond to consecutive distance bins of a 30 deg x 30 deg region around the Orion nebula; each panel shows a differential extinction map (i.e. the amount of extinction that is added in the respective distance range). We thus observe how the extinction from the molecular dust clouds gradually fills the Galactic plane with increasing distance. This effect is in principle already observable in the median {\tt StarHorse} {\it Gaia} DR2 extinction maps of the Orion region shown in Fig. 11 of \citet{Anders2019}, but now the scatter in extinction per star (and thus also per pixel) is lower and the amount of unphysical artefacts has diminished. 

It is thus possible to infer the 3D dust distribution almost by eye from sequential 2D slices such as those shown in Fig. \ref{fig:orion}. In the case of farther regions and/or greater statistical noise, however, the inversion problem of inferring the 3D dust distribution in the extended solar neighbourhood from photo-astrometric data requires significant conceptional and computational effort \citep[e.g.]{Lallement2014, Lallement2019, Lallement2022, Sale2018, Green2019, Leike2020, RezaeiKh.2020}.

\subsection{Metal-poor stars}\label{sec:vmp}

\begin{figure}\centering    \includegraphics[width=0.49\textwidth]{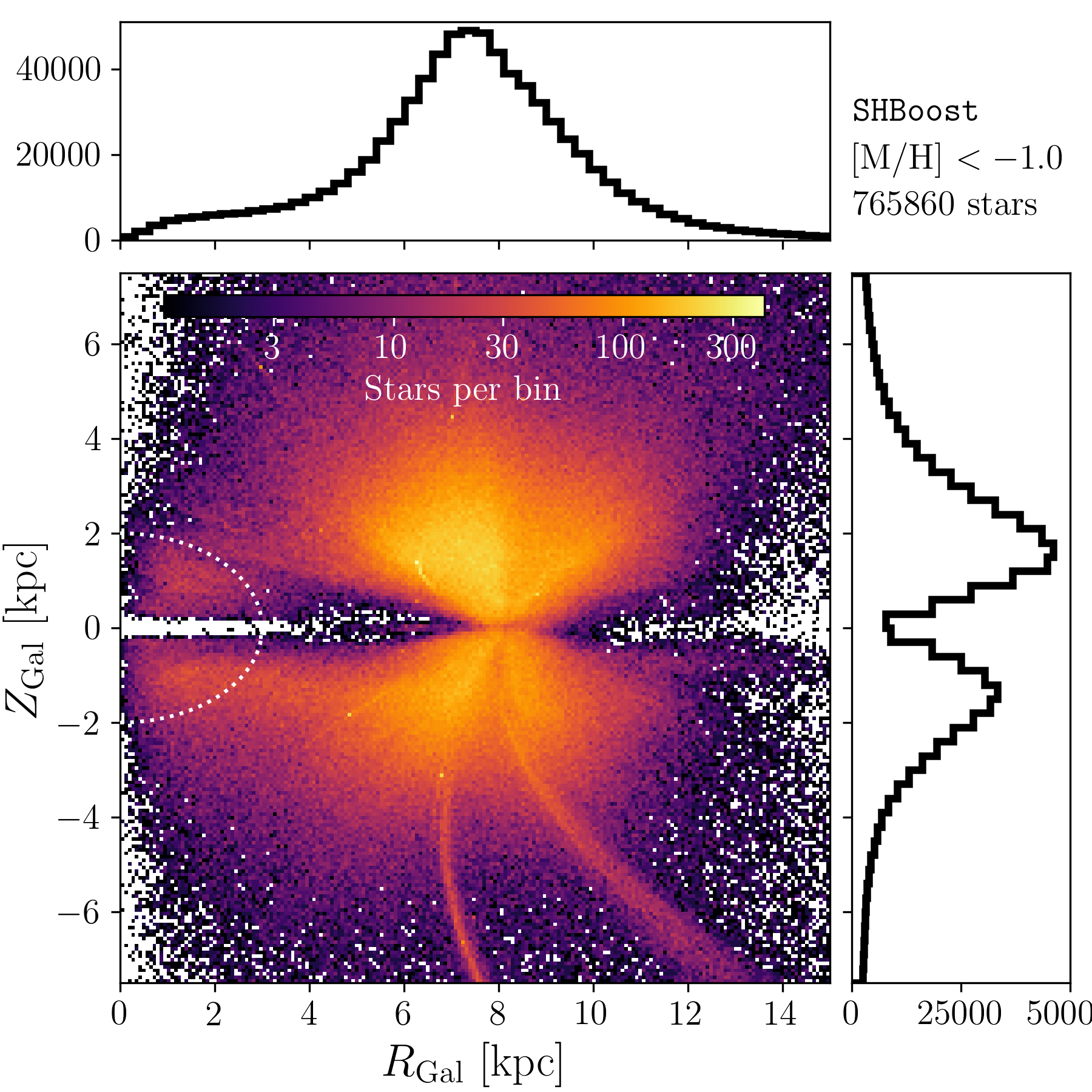}
        \caption{$Z_{\rm Gal}$ vs $R_{\rm Gal}$ map of low-metallicity candidate stars (defined as stars that have {\tt xgbdist\_met\_mean} $<-1$ and {\tt xgb\_met\_outputflag}$=="0"$). There are 766k stars in the region covered by the map (943k in total). The white dashed ellipse marks the typical extent of a classical bulge region.}
        \label{fig:vmp_maps}
\end{figure}

The plethora of recent works providing improved stellar metallicity estimates from {\it Gaia} XP spectra (e.g. \citealt{Andrae2023, Zhang2023, Xylakis-Dornbusch2024, Yao2024}) is (to some degree) driven by the need for reliable input catalogues of ongoing and upcoming surveys targeting metal-poor and very metal-poor stars \citep[e.g.][]{Christlieb2019, Conroy2019, Cioni2019, Chiappini2019, Arentsen2020}. 

Pre-{\it Gaia} DR3 surveys of metal-poor stars, such as SkyMapper \citep{Keller2007}, J-PAS/J-PLUS \citep{Marin-Franch2012}, Pristine \citep{Starkenburg2017}, or PIGS \citep{Arentsen2019}, have  fundamentally relied on the combination of narrow- and intermediate-band photometry to select candidates (e.g. \citealt{Frebel2015, Youakim2017, Whitten2019, Chiti2021, Galarza2022}). The high signal-to-noise ratios of the {\it Gaia} XP spectra make it possible to construct synthetic narrow-band filters from these spectra, for example using the {\tt gaia-xpy} package \citep{Ruz-Mieres2022}, without the need for costly narrow-band observations with ground-based telescopes.

In this work, we have used the {\it Gaia} DR3 XP spectra in their native form (using the coefficients of the Hermite basis functions that describe the spectra; \citealt{Carrasco2021}). Other works, such as \citet{Andrae2023}, have also used the synthetic narrow-band magnitudes and focussed on a narrower portion of the Hertzsprung-Russell diagram. This enabled their study to achieve a higher formal precision in the obtained metallicities (see App. \ref{sec:andrae2023}).

The comparison with open and globular clusters in Appendix \ref{sec:feh_calib} shows that our {\tt xgb\_met} global metallicity estimates are systematically overestimated for low metallicities (see also Fig. \ref{fig:validation-onetoone} and Appendix \ref{sec:andrae2023}). However, these estimates are still useful to select a pure sample of low-metallicity candidates to be followed up with spectroscopic surveys. Figure \ref{fig:vmp_maps} shows an example selection of bona-fide metal-poor stars (selected as {\tt xgbdist\_met\_mean} +  {\tt xgbdist\_met\_std} $<-1$) closer than 5 kpc to the Galactic plane. A comparison with Fig. \ref{fig:feh_calib} shows that while this selection is not optimised for very or extremely metal-poor stars (see \citealt{Xylakis-Dornbusch2024} or \citealt{Yao2024} for dedicated attempts), it does result in a disc sample that has a purity of metal-poor stars (i.e. spectroscopic [M/H] $< -1$) greater than 90\%. The metallicity range $-2 \lesssim {\rm [M/H]} \lesssim -1$ is in itself a very interesting regime that provides crucial hints to deciphering the early evolution and merger history of the Milky Way's disc \citep[e.g.][]{Nepal2024}. The selection shown in Fig. \ref{fig:vmp_maps} was therefore chosen as the base sample for the 4MIDABLE-LR \citep{Chiappini2019} sub-survey on metal-poor disc stars, which will commence observations as part of 4MOST project \citep{deJong2019} in 2025 \citep{deJong2022}.

\subsection{Super-metal-rich stars}

\begin{figure}\centering
        \includegraphics[width=0.49\textwidth]{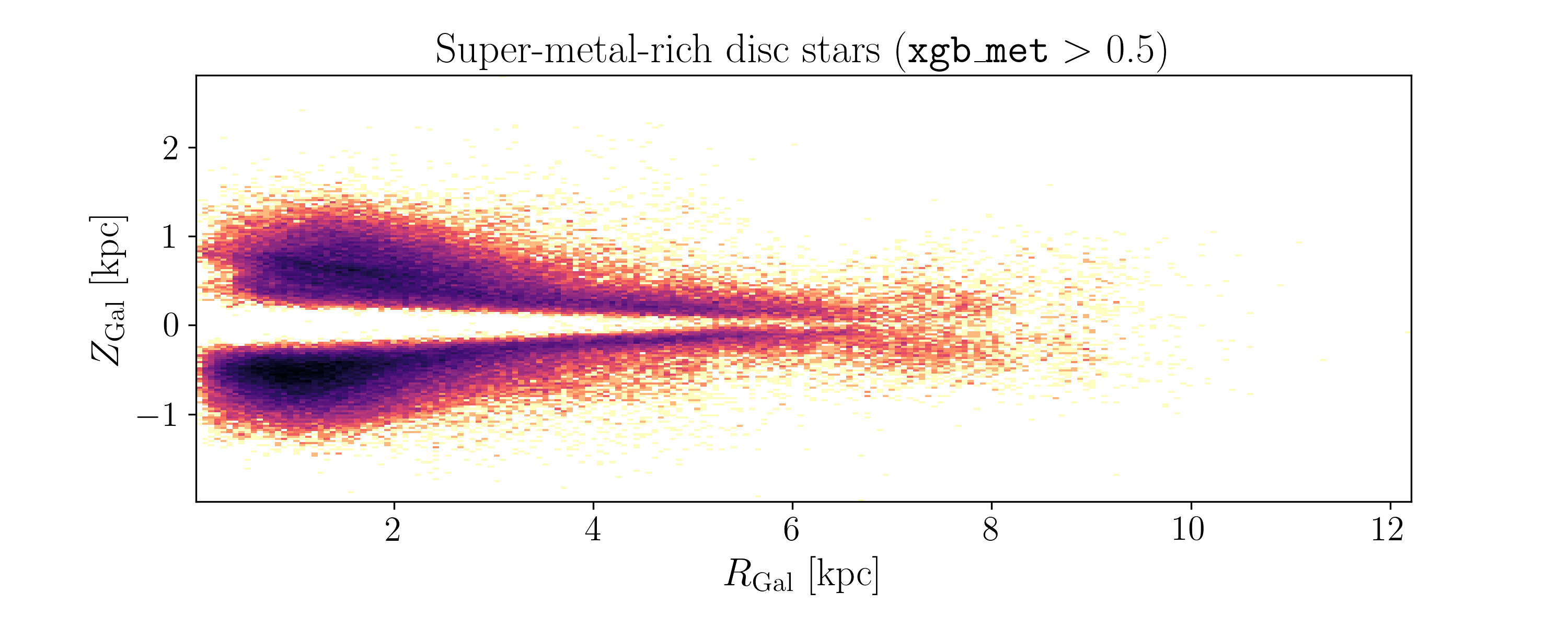}
        \includegraphics[width=0.49\textwidth, trim={0 0 0 2cm},clip]{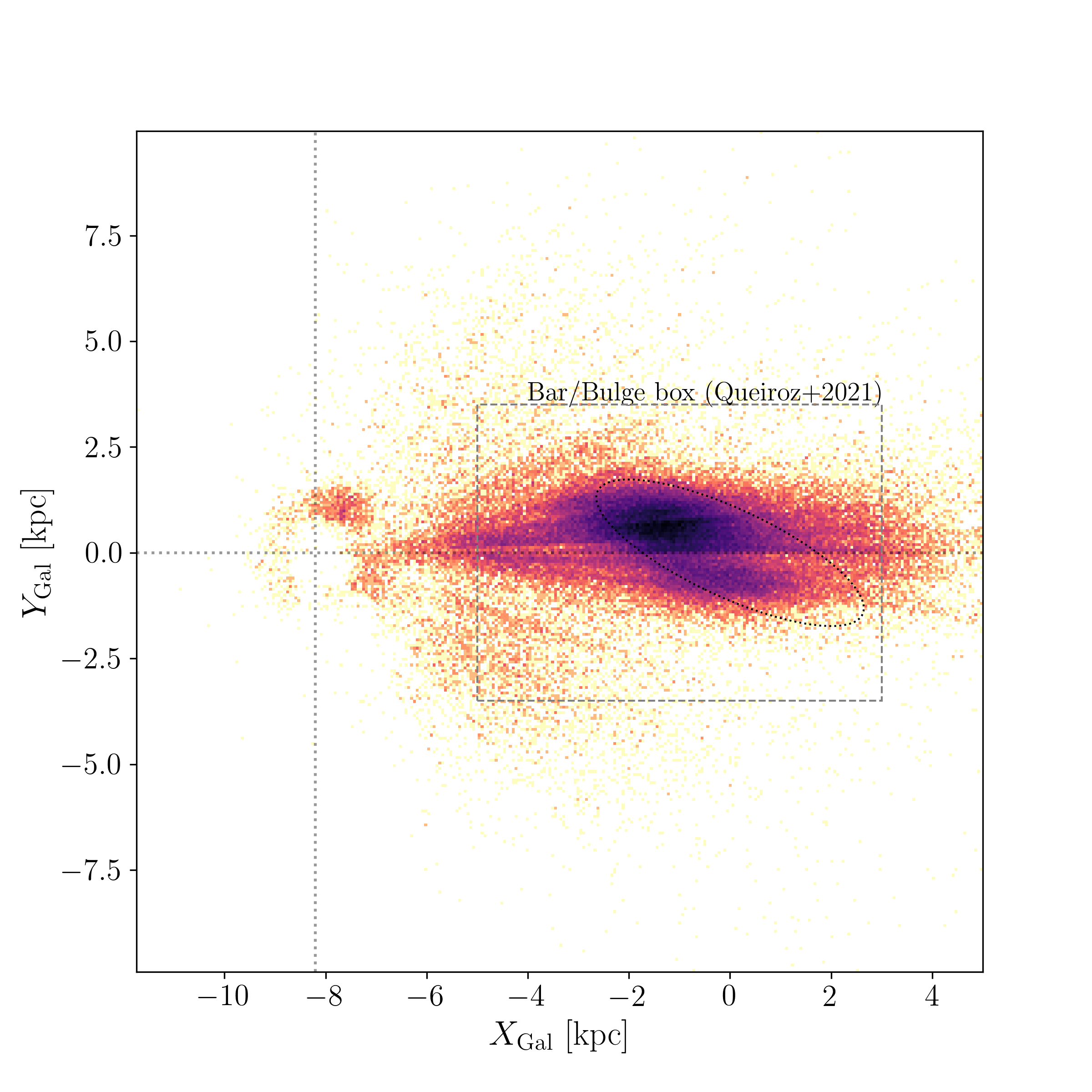}
        \caption{Galactic maps of super-metal-rich stars ({\tt xgbdist\_met} $>+0.5$). The lower panel shows a top-down view of the Galaxy, marking the solar position (dotted lines), the approximate extent of the Galactic bar (ellipse), and the bulge and bar region studied by \citet{Queiroz2021}.}
        \label{fig:smr_maps}
\end{figure}

As shown in Appendix \ref{sec:literature}, our metallicity estimates are most reliable in the solar and super-solar regime. An immediate application is thus the mapping of super-metal-rich (SMR) stars. Figure \ref{fig:smr_maps} shows the distribution of stars with {\tt xgbdist\_met} $>+0.5$ in the Galaxy, highlighting that most of these stars are located in the Galactic bar and bulge region (which is known to host a rich reservoir of such stars; see e.g. \citealt{Queiroz2021, Rix2024}). It also shows a few artefacts that are likely to be related to erroneous distances (in particular, the ring-like structure at $\sim 1$ kpc from the Sun; see also \citealt{Anders2019}).

\begin{figure}\centering
        \includegraphics[width=0.49\textwidth]{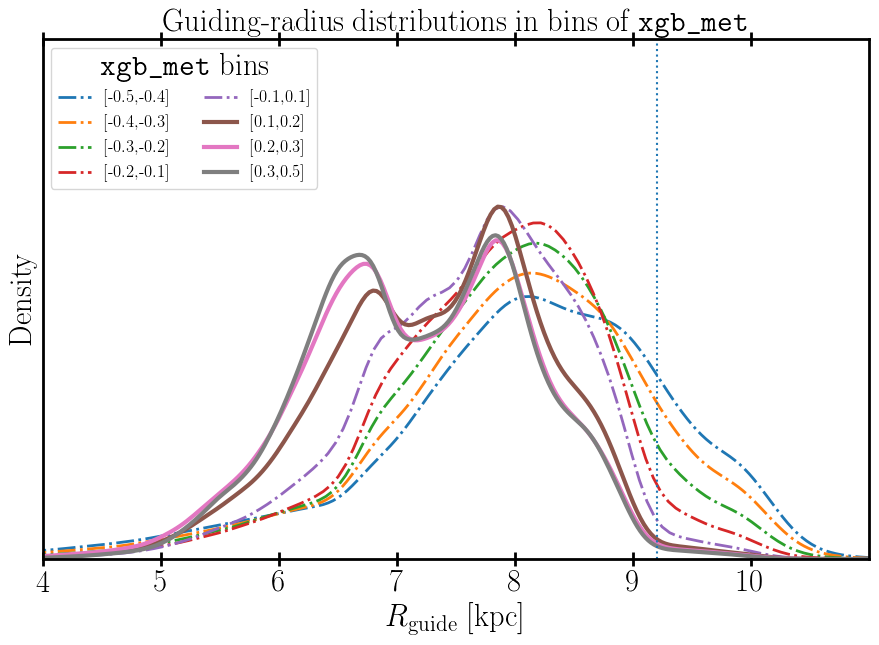}
        \caption{Guiding-radius distributions of the XP sample with radial velocities from {\it Gaia} RVS, in bins of {\tt xgbdist\_met}. The curves of SMR stars are highlighted as thicker lines. The dotted vertical line at $R_{\rm guide}$ highlights the point after which the density of metal-rich stars reaches a floor, which might possibly be related to the outer Lindblad resonance of the Galactic bar.}
        \label{fig:rguide_metbins}
\end{figure}

Figure \ref{fig:rguide_metbins} demonstrates another possible application of our catalogue: the slicing of the XP sample into metallicity bins, which allows for statistical stellar-population studies that are otherwise difficult to carry out. In particular, Fig. \ref{fig:rguide_metbins} shows the distribution of orbital guiding radii (calculated assuming a flat rotation curve, roughly equivalent to angular momentum) in bins of metallicity between $-0.5$ and $+0.5$ for over 7 million stars in the disc ($|Z_{\rm Gal}| < 1$ kpc and heliocentric distances $\lesssim 4$ kpc). The most obvious observation is of course the gradual inward shift of the $R_{\rm guide}$ distributions with growing metallicity, caused by the disc's negative radial abundance gradient (e.g. \citealt{Janes1979, Grenon1987, Anders2014, Luck2018}). We also observe a second effect: the super-solar metallicity stars ({\tt xgb\_met}$>+0.1$) show peaks in the $R_{\rm guide}$ distribution similar to the bi-modality, which was found and discussed in \citet{Nepal2024a} using the much smaller {\it Gaia} DR3 RVS CNN dataset (\citealt{Guiglion2024}, containing $\sim 169\,000$ MSTO stars). The XP sample, thanks to its larger statistics, now allows us to also confirm another signature that was already hinted at by the data shown in \citet{Nepal2024a}: There is an abrupt scarcity of metal-rich stars ({\tt xgb\_met}$>+0.1$) beyond $\sim 9.2$ kpc (see also \citealt{Khoperskov2022}). It has been suggested that the Galactic bar's outer Lindblad resonance (OLR) may limit the migration of stars from the inner galaxy beyond this location (see e.g.  \citealt{Halle2015, Khoperskov2020}). We therefore suggest that this break in density at $\sim 9.2$ kpc could mark the position of the bar's OLR, which provides a natural barrier for the radial migration of stars on cold orbits. 

These trends displayed by the metal-rich stars, previously shown on the basis of smaller samples, is impossible to explain only by selection biases and hint at a major role of the galactic bar in the chemo-dynamical evolution of the disc. In a future work (in preparation), we aim to use the SMR stars from the {\tt SHBoost} catalogue to constrain the bar properties.

\section{Conclusions and lessons for future work}\label{sec:conclusions}

In this paper, we present the {\tt SHBoost} catalogue, derived from a set of {\tt xgboost} regression models that determine stellar parameters and line-of-sight extinctions for more than 217 million stars with {\it Gaia} DR3 XP spectra. Through the curation of a high-quality training set (mostly coming from the spectroscopic survey data compiled in \citealt{Queiroz2023}), the code yields results that are competitive with its classical isochrone-fitting counter-part, the {\tt StarHorse} {\it Gaia} EDR3 data presented in \citet{Anders2022}. For reference, Table \ref{tab:summary} summarises the stellar-parameter, distance, and extinction catalogues available from our group.

\begin{table*}
    \caption{Stellar-parameter catalogues produced by the {\tt StarHorse} group over the past years.}
    \begin{tabular}{llcccccccc}
    Reference & Scope & $N_{\rm stars}$ & $d$ & $A_V$ & $T_{\rm eff}$ & $\log g$ & [M/H] & Mass & Age \\
    \hline
    \citet{Queiroz2018}  &  4 spectroscopic surveys & 0.5M & x & x & x & x & x & \\
    \citet{Anders2018}  &  HARPS-GTO sample & 0.001M & x & x & x & x & x & x & x\\
    \citet{Anders2019}  &  {\it Gaia} DR2 + 2MASS+AllWISE ($G<18$) & 266M & x & x & x & x & x & x \\
    \citet{Queiroz2020}  &  5 spectroscopic surveys & 1.6M & x & x & x & x & x & x & \\
    \citet{Anders2022}  &  {\it Gaia} EDR3 + photometry ($G<18.5$) & 362M & x & x & x & x & x & x & x\tablefootmark{*}\\
    \citet{Queiroz2023}  &  8 spectroscopic surveys & 11.1M & x & x & x & x & x & x & x \\
    Nepal et al. (in prep.)  &  {\it Gaia} DR3 RVS CNN & 0.8M & x & x & x & x & x & x & x \\
    \hline
    This work  &  {\it Gaia} DR3 XP spectra & 217M &  & x & x & x & x & x \\
    \end{tabular}
    \tablefoot{\tablefoottext{*}{Age estimates for the {\tt StarHorse} EDR3 run can only be obtained from the approximate joint posterior PDFs published at \url{https://data.aip.de/projects/starhorse2021.html}.}}
    \label{tab:summary}
\end{table*}

Our new {\tt SHBoost} catalogue comes with reliable uncertainties estimated via {\tt xgboost-distributions}. We achieve median (best 25\% percentile in brackets) precisions of 0.20 (0.11) mag in line-of-sight extinction, $A_V$, 0.014 (0.010) dex for effective temperature, $\log T_{\rm eff}$ (or 174 [119] K in $T_{\rm eff}$), 0.20 (0.13) mag in surface gravity, $\log g$, 0.18 (0.14) dex in metallicty [M/H], and 12\% (8\%) in mass. The validation of the individual uncertainty estimates shows that they are correctly estimated. In addition, our {\tt xgboost} models can be interpreted by using SHAP values (see Sect. \ref{sec:shap} and Fig. \ref{fig:shap}). 

We demonstrate that the combination of training data from several different surveys produces satisfactory results in parameter regimes that are traditionally not well covered by large-scale spectroscopic surveys (e.g. white dwarfs, OB stars, sdO stars) and other catalogues based on {\it Gaia} XP spectra. Other available stellar-parameter catalogues derived from XP spectra mostly rely on APOGEE data (e.g. \citealt{Andrae2023, Zhang2023, Fallows2024}).

The whole inference pipeline (building the training set, training of the {\tt xgboost} regression of stellar parameters, post-processing) could be run in 2 days on a 48-core machine at the Leibniz-Institut für Astrophysik Potsdam. This is a significant improvement in computational speed compared to the latest {\tt StarHorse} EDR3 run, which took roughly three weeks to complete on a similar machine. We emphasise, however, that the need for classical Bayesian inference (at least for the construction of large robust training sets) will persist in the future. 

In Sect. \ref{sec:results}, we showcase several potential applications of our new catalogue, including differential extinction maps, metallicity trends in the Milky Way, and extended maps of young massive stars, metal-poor stars, and metal-rich stars. In particular, our hot star map reveals a new stellar void adjacent to the Vela region, suggesting that more such inhomogeneities are hiding in the extended solar neighbourhood.

The distribution of metal-poor stars shows homogeneity around the Sun and a spherical over-density in the Galactic centre, corroborating previous findings of an old bulge population distinct from the boxy peanut bulge \citep{Barbuy2018, Zoccali2019, Rix2022, Arentsen2024}. In addition, mapping metal-rich stars across the Galaxy reveals a significant reservoir of these stars in the inner regions \citep{Queiroz2021, Rix2024}. The large number of metal-rich stars in our catalogue provides new opportunities to study the formation and evolution of the inner disk and bar structure. As recently found in \citet{Nepal2024a}, but now with a much larger number of stars, we have identified a bi-modality in the guiding radius distribution of super-metal-rich stars and a scarcity of such stars beyond a Galactocentric distance of $\sim$9.2 kpc, which may be related to the outer Lindblad resonance of the Galactic bar.

In view of {\it Gaia} DR4, which will contain a much greater data volume (including many millions of previously unavailable XP and RVS spectra), we foresee that convolutional neural-network approaches will likely prevail over future tree-based implementations. This is supported by recent developments, which allow for similar levels of performance in terms of precision and computational cost, while naturally allowing for the inclusion of uncertain and correlated output labels \citep[e.g.][]{Fallows2024}. 

\section*{Data availability}

Table \ref{tab:datamodel1} provides the data model for the provided {\tt SHBoost} output tables, available at the CDS. Alternatively, the data can also be accessed via a dedicated web page\footnote{\url{https://data.aip.de/projects/shboost2024.html}}, where we provide detailed instructions and python notebooks that illustrate how to download and use the catalogue efficiently.

\begin{acknowledgements}
This work was partially funded by the Spanish MICIN/AEI/10.13039/501100011033 and by the "ERDF A way of making Europe" funds by the European Union through grants RTI2018-095076-B-C21, PID2021-122842OB-C21, PID2021-125451NA-I00, and CNS2022-135232, and the Institute of Cosmos Sciences University of Barcelona (ICCUB, Unidad de Excelencia ’Mar\'{\i}a de Maeztu’) through grant CEX2019-000918-M. FA acknowledges financial support from MCIN/AEI/10.13039/501100011033 through grants IJC2019-04862-I and RYC2021-031638-I (the latter co-funded by European Union NextGenerationEU/PRTR). GG acknowledges support by Deutsche Forschungsgemeinschaft (DFG, German Research Foundation) – project-IDs: eBer-22-59652 (GU 2240/1-1 "Galactic Archaeology with Convolutional Neural-Networks: Realising the potential of Gaia and 4MOST"). This project has received funding from the European Research Council (ERC) under the European Union’s Horizon 2020 research and innovation programme (Grant agreement No. 949173).\\

This work made use of the following software packages: \texttt{astropy} \citep{astropy:2013, astropy:2018, astropy:2022}, \texttt{Jupyter} \citep{2007CSE.....9c..21P, kluyver2016jupyter}, \texttt{matplotlib} \citep{Hunter:2007}, \texttt{numpy} \citep{numpy}, \texttt{pandas} \citep{mckinney-proc-scipy-2010, pandas_10426137}, \texttt{python} \citep{python}, \texttt{scipy} \citep{2020SciPy-NMeth, scipy_10909890}, \texttt{astroquery} \citep{2019AJ....157...98G, astroquery_10799414}, \texttt{Cython} \citep{cython:2011}, \texttt{h5py} \citep{collette_python_hdf5_2014, h5py_7560547}, \texttt{scikit-learn} \citep{scikit-learn, sklearn_api, scikit-learn_11237090} and \texttt{seaborn} \citep{Waskom2021}.
This research has made use of NASA's Astrophysics Data System.
Some of the results in this paper have been derived using \texttt{healpy} and the HEALPix package\footnote{\url{http://healpix.sourceforge.net}} \citep{Zonca2019, 2005ApJ...622..759G, healpy_11337740}.
Software citation information were aggregated using \texttt{\href{https://www.tomwagg.com/software-citation-station/}{The Software Citation Station}} \citep{software-citation-station-paper, software-citation-station-zenodo}.\\

This work has made use of data from the European Space Agency (ESA) mission {\it Gaia} (\url{http://www.cosmos.esa.int/gaia}), processed by the {\it Gaia} Data Processing and Analysis Consortium (DPAC,
\url{http://www.cosmos.esa.int/web/gaia/dpac/consortium}). Funding for the DPAC has been provided by national institutions, in particular the institutions participating in the {\it Gaia} Multilateral Agreement.  

Funding for the SDSS Brazilian Participation Group has been provided by the Minist\'erio de Ci\^encia e Tecnologia (MCT), Funda\c{c}\~ao Carlos Chagas Filho de Amparo \`a Pesquisa do Estado do Rio de Janeiro (FAPERJ), Conselho Nacional de Desenvolvimento Cient\'{\i}fico e Tecnol\'ogico (CNPq), and Financiadora de Estudos e Projetos (FINEP). Funding for the Sloan Digital Sky Survey IV has been provided by the Alfred P. Sloan Foundation, the U.S. Department of Energy Office of Science, and the Participating Institutions. SDSS-IV acknowledges support and resources from the Center for High-Performance Computing at the University of Utah. The SDSS web site is \url{www.sdss.org}. SDSS-IV is managed by the Astrophysical Research Consortium for the Participating Institutions of the SDSS Collaboration including the Brazilian Participation Group, the Carnegie Institution for Science, Carnegie Mellon University, the Chilean Participation Group, the French Participation Group, Harvard-Smithsonian Center for Astrophysics, Instituto de Astrof\'isica de Canarias, The Johns Hopkins University, 
Kavli Institute for the Physics and Mathematics of the Universe (IPMU) / University of Tokyo, Lawrence Berkeley National Laboratory, Leibniz-Institut f\"ur Astrophysik Potsdam (AIP), Max-Planck-Institut f\"ur Astronomie (MPIA Heidelberg), Max-Planck-Institut f\"ur Astrophysik (MPA Garching), Max-Planck-Institut f\"ur Extraterrestrische Physik (MPE), National Astronomical Observatory of China, New Mexico State University, New York University, University of Notre Dame, Observat\'ario Nacional / MCTI, The Ohio State University, Pennsylvania State University, Shanghai Astronomical Observatory, United Kingdom Participation Group, Universidad Nacional Aut\'onoma de M\'exico, University of Arizona, University of Colorado Boulder, University of Oxford, University of Portsmouth, University of Utah, University of Virginia, University of Washington, University of Wisconsin, Vanderbilt University, and Yale University.

Guoshoujing Telescope (the Large Sky Area Multi-Object Fiber Spectroscopic Telescope LAMOST) is a National Major Scientific Project built by the Chinese Academy of Sciences. Funding for the project has been provided by the National Development and Reform Commission. LAMOST is operated and managed by the National Astronomical Observatories, Chinese Academy of Sciences.

Funding for RAVE has been provided by: the Australian Astronomical Observatory; the Leibniz-Institut f\"ur Astrophysik Potsdam (AIP); the Australian National University; the Australian Research Council; the French National Research Agency; the German Research Foundation (SPP 1177 and SFB 881); the European Research Council (ERC-StG 240271 Galactica); the Istituto Nazionale di Astrofisica at Padova; The Johns Hopkins University; the National Science Foundation of the USA (AST-0908326); the W. M. Keck foundation; the Macquarie University; the Netherlands Research School for Astronomy; the Natural Sciences and Engineering Research Council of Canada; the Slovenian Research Agency; the Swiss National Science Foundation; the Science \& Technology Facilities Council of the UK; Opticon; Strasbourg Observatory; and the Universities of Groningen, Heidelberg and Sydney. The RAVE web site is at \url{https://www.rave-survey.org}.

This work has also made use of data from {\it Gaia}-ESO based on data products from observations made with ESO Telescopes at the La Silla Paranal Observatory under programme ID 188.B-3002.\\

\end{acknowledgements}

\bibliographystyle{aa}
\bibliography{shboost_final}

\appendix

\section{Comparison to previous catalogues}\label{sec:literature}

\begin{figure*}\centering
        \includegraphics[width=0.95\textwidth]{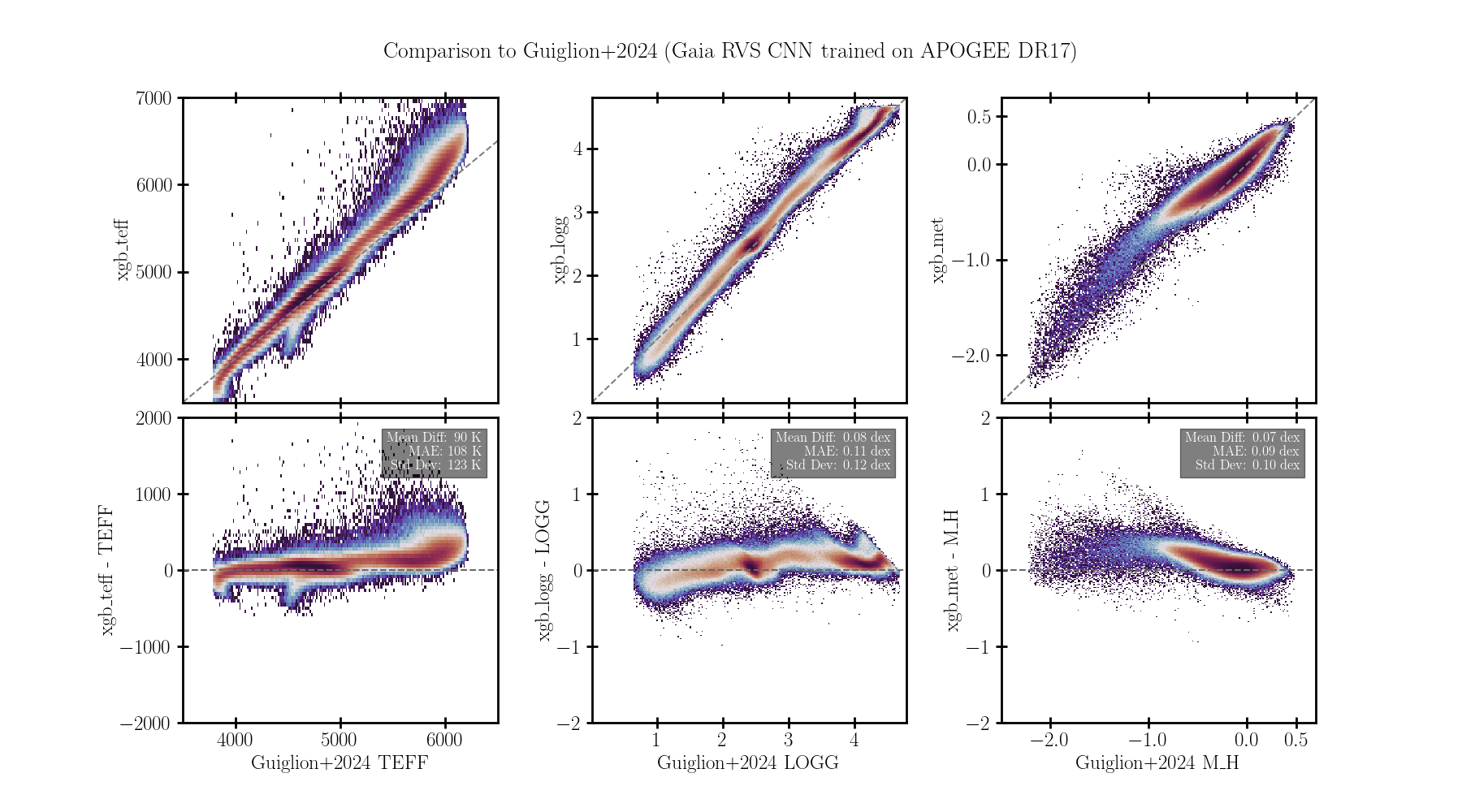}
        \caption{Comparison to the stellar-parameter catalogue of \citet{Guiglion2024}, based on {\it Gaia} DR3 data (especially RVS spectra, but also XP spectra, parallaxes, and broad-band photometry). Their catalogue was obtained with a convolutional neural network; the labels were trained on APOGEE DR17.}
        \label{fig:guiglion}
\end{figure*}

An exhaustive comparison to similar efforts in the literature is beyond the scope of this paper; the comparisons shown in this appendix serve as illustrations of the magnitude of systematic differences one can expect between the various stellar-parameter catalogues based on {\it Gaia} DR3 that have become available in the past four years.

\subsection{{\it Gaia} DR3 RVS CNN catalogue \citep{Guiglion2024}} \label{sec:gui2024}

In Fig. \ref{fig:guiglion} we compare the {\tt SHBoost} parameterisation with the \emph{Gaia} RVS atmospheric parameters derived by \citet{Guiglion2024} based on a hybrid Convolutional Neural-Network (CNN). In their study, \citet{Guiglion2024} combined the full \emph{Gaia} DR3 data products in the form of {\it Gaia} DR3 RVS spectra, parallaxes, $G, G_{BP},$ and $G_{RP}$ magnitudes, as well as XP coefficients. This unique combination provides precise atmospheric parameters and robust [$\alpha$/M] estimates even for low signal-to-noise RVS spectra (down to S/N$_{\rm RVS}=15$), for about 800\,000 stars. The authors characterised the [$\alpha$/M] vs [M/H] bi-modality from the inner to the  outer disc, for the first time with \emph{Gaia} RVS data. The CNN was trained on a high-quality training sample of $\sim40\,000$ stars in common between the \emph{Gaia} DR3 RVS sample and APOGEE DR17. We cross-matched the RVS-CNN catalogue of \citet{Guiglion2024} with {\tt SHBoost} catalogue. After the recommended quality cuts for the RVS-CNN catalogue (see \citet{Guiglion2024} for details) and requiring \text{xgb\_inputflag == 'XP5pGBPRPJHKsW1W2W3W4'} for the {\tt SHBoost} catalogue, we obtained a total of 655\,950 stars. 

\begin{figure*}\centering
        \includegraphics[width=0.8\textwidth]{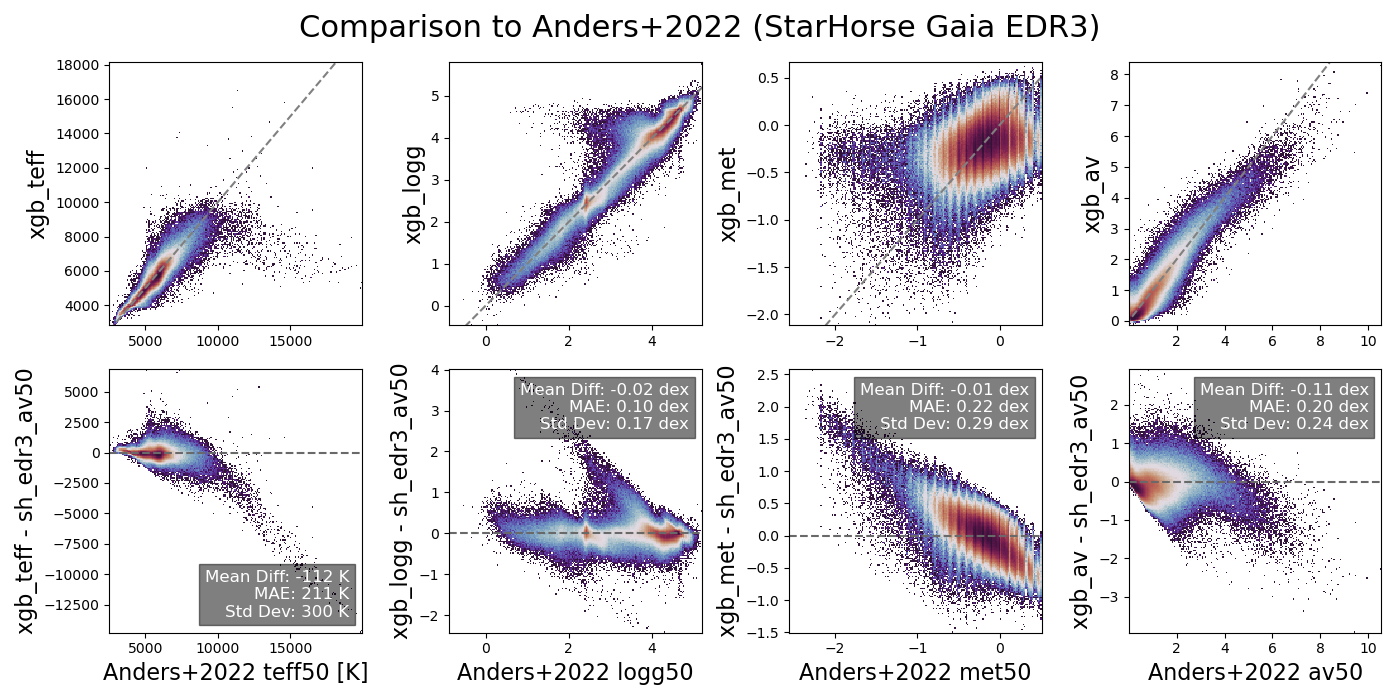}
        \caption{Comparison of the {\tt SHBoost} parameters with the {\tt StarHorse} {\it Gaia} EDR3 parameters \citep{Anders2022}. Top panels: One-to-one comparisons of effective temperature, surface gravity, metallicity, and extinction (from left to right). Bottom panels: Residuals.}
        \label{fig:sh}
\end{figure*}

As can be appreciated in Fig. \ref{fig:guiglion}, we obtain very similar results to \citet{Guiglion2024} for the three stellar parameters in common ($T_{\rm eff}, \log g$, and [M/H]), modulo some small systematic trends. This is reassuring, especially since \citet{Guiglion2024} used also the {\it Gaia} RVS spectra (with spectral resolution $R\sim 11\,000$), which add much more precise information on metallicity than can be obtained from the low-resolution XP spectra. 

By construction, our allowed range of $T_{\rm eff}$ is larger than in the catalogue of \citet{Guiglion2024}, which used APOGEE DR17 as a training set. This explains the deviating behaviour at the edges of the $T_{\rm eff}$ range shown in the left panels of Fig. \ref{fig:guiglion}. The $\log g$ comparison (middle panels) shows systematic differences below the 0.2 dex level (mean difference with respect to \citealt{Guiglion2024} is +0.08 dex). For metallicity (right panels of Fig. \ref{fig:guiglion}), we find a very good agreement in the disc-like metallicity regime ($> -0.6$ dex), while at lower metallicities our values are systematically higher than the ones of \citet{Guiglion2024}, suggesting that our metallicities are overestimated in this regime (see also the sections below).
We also inspected the weak horizontal branch visible in the top right panel of Fig. \ref{fig:guiglion} and find that for these stars our metallicities agree very well with the high-resolution metallicity scale of APOGEE, suggesting that for this sample the \citet{Guiglion2024} metallicities are underestimated. 

\subsection{{\it Gaia} EDR3 {\tt StarHorse} \citep{Anders2022}} \label{sec:sh2021}

In Fig. \ref{fig:sh} we compare our {\tt SHBoost} results with the results obtained from traditional Bayesian isochrone fitting to photo-astrometric data with {\tt StarHorse} \citep{Anders2022}. In that paper we published a catalogue of distances, extinctions, and stellar parameters for 362 million {\it Gaia} EDR3 stars brighter than $G<18.5$ (see Table \ref{tab:summary}).

We observe that the effective temperature estimates between both works agree well for the vast majority of the data (between 3000 K and 10\,000 K), albeit with a considerable scatter (MAE of 211 K). At greater $T_{\rm eff}$, we had already shown in \citet{Anders2022} that the {\tt StarHorse} EDR3 $T_{\rm eff}$ scale is not very reliable, probably due to the sparse sampling of massive-star isochrones, binarity, and the initial-mass-function prior.
We also refer to the discussion of hot luminous stars in Sect. \ref{sec:hot}.

For $\log g$ and $A_V$ (second and forth panel of Fig. \ref{fig:sh}, respectively), we see a very good concordance between {\tt StarHorse} EDR3 and {\tt SHBoost}, with little global shifts ($-0.02$ dex and $-0.11$ mag, respectively) and small scatter around the one-to-one correspondence (MAEs of $0.10$ dex and $0.20$ mag, respectively). The group of stars with {\tt StarHorse} $\log g\simeq3-4$ which has {\tt xgb\_logg} around 4.5--5 seems to be composed of genuine dwarf stars, which were affected by uncertain {\it Gaia} (E)DR3 parallax measurements, which lead to a {\tt StarHorse} solution that is compatible for stars in either the dwarf, sub-giant, or giant evolutionary stages. 

\begin{figure*}\sidecaption
        \includegraphics[width=0.65\textwidth]{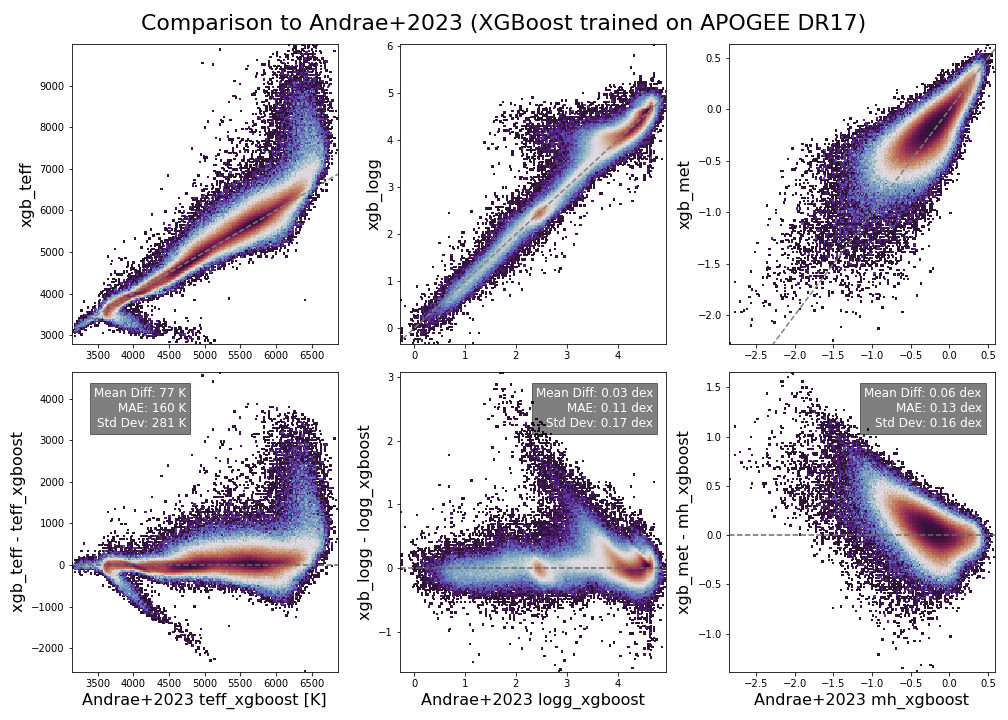}
        \caption{Comparison to the stellar-parameter catalogue of \citet{Andrae2023}, trained mostly on APOGEE DR17.}
        \label{fig:andrae}
\end{figure*}

The greatest discrepancy of our {\tt SHBoost} catalogue with the {\tt StarHorse} EDR3 results arises for the metallicities (third column of Fig. \ref{fig:sh}). This is generally expected because the {\tt StarHorse} metallicity estimates are only based on broad-band photometry (in the best cases including $griz$ photometry from PanSTARRS-1 or SkyMapper); whereas the {\tt SHBoost} results use the low-resolution {\it Gaia} XP spectra, which contain much more precise information about both narrow and broad absorption features \citep{Weiler2023}. The large scatter with respect to the {\tt StarHorse} EDR3 metallicity scale is thus fully anticipated. The systematic trend at low metallicity, however, is less intuitive. Since our {\tt SHBoost} metallicities agree well with the ones obtained by \citet{Guiglion2024}, as shown in Sect. \ref{sec:gui2024}, we suggest that the {\tt StarHorse} EDR3 photo-astrometric metallicities should be used with great caution, and not on an individual-star level.

\subsection{{\it Gaia} DR3 XP spectra + {\tt xgboost}: \citet{Andrae2023}} \label{sec:andrae2023}

The first published stellar parameters derived from {\it Gaia} XP spectra using {\tt xgboost} regression were delivered by \citet{Andrae2023}. These authors also included as input columns, in addition to the XP coefficients, synthetic narrow-band photometry derived from the XP spectra, which allowed them to obtain more precise metallicity estimates.

In Fig. \ref{fig:andrae} we compare our results to the catalogue of \citet{Andrae2023}. Some of the patterns in the $T_{\rm eff}$ and $\log g$ comparisons (left and middle column, respectively) are similar to the ones seen in Figs. \ref{fig:guiglion} and \ref{fig:sh}. The larger range of effective temperatures in our training set allows for a better coverage of especially upper main-sequence stars. The diagonal branch in the lower left panel of Fig. \ref{fig:andrae} is due to a set of stars potentially misclassified in the \citet{Andrae2023} catalogue. 
The $\log g$ comparison in the middle panels of Fig. \ref{fig:andrae} shows little dispersion (MAE: 0.11 dex) and only small offsets (mean: +0.03 dex). We also see similar trends of the $\log g$ systematics with respect to both the \citet{Andrae2023} and \citet{Anders2022} catalogues.

The most discrepant results with respect to \citet{Andrae2023} arise for metallicity. On the one hand, this is due to the different metallicity definitions used (\citealt{Andrae2023} trained on the APOGEE DR17 [Fe/H] labels, while we use [M/H], which results in significant offsets for [$\alpha$/Fe]-enhanced populations; see also Sect. \ref{sec:feh_calib}). On the other hand, the training set of \citet{Andrae2023} itself is much smaller, but less heterogeneous than ours, resulting in better performance at least for $G<16$, which is the range well covered by their training data. In addition, it appears that their approach to include also synthetic photometry in the training data further improved the precision. In principle, {\tt xgboost} is able to fit also complex non-linear dependencies of multiple input quantities accurately, but the work of \citet{Andrae2023} shows that in particular for metallicity it does help to include more explicit information about metallicity-sensitive features (such as the synthetic Pristine CaHK filter magnitude) in the training data. 

\section{Data model}\label{sec:datamodel}

Table \ref{tab:datamodel1} provides the data model for the provided {\tt SHBoost} output tables, available at the CDS and the AIP web page\footnote{\url{https://data.aip.de/projects/shboost2024.html}}, or through its DOI\footnote{\url{https://doi.org/10.17876/data/2024_3}}.

\begin{table*}
\centering
\caption{Data model of the {\it Gaia} DR3 {\tt SHBoost} catalogue.}
\begin{tabular}{rlll}
ID & Column name & Unit & Description \\
\hline
0 & {\tt source\_id} &  & {\it Gaia} DR3 unique source identifier \\ 
1 & {\tt xgb\_av} & mag & line-of-sight extinction at $\lambda=5420\, \AA$, $A_{\rm V}$, {\tt xgboost} point estimate \\ 
2 & {\tt xgbdist\_av\_mean} & mag & line-of-sight extinction at $\lambda=5420\, \AA$, $A_{\rm V}$, {\tt xgboost-distribution} mean value  \\ 
3 & {\tt xgbdist\_av\_std} & mag & line-of-sight extinction at $\lambda=5420\, \AA$, $A_{\rm V}$, {\tt xgboost-distribution} standard deviation \\ 
4 & {\tt xgb\_logteff} & [K] & effective temperature, {\tt xgboost} point estimate \\ 
5 & {\tt xgb\_logteff\_mean} & [K] & effective temperature, {\tt xgboost-distribution} mean value \\ 
6 & {\tt xgb\_logteff\_std} & [K] & effective temperature, {\tt xgboost-distribution} standard deviation \\ 
7 & {\tt xgb\_logg} & [cgs] & surface gravity, {\tt xgboost} point estimate \\ 
8 & {\tt xgb\_logg\_mean} & [cgs] & surface gravity, {\tt xgboost-distribution} mean value \\ 
9 & {\tt xgb\_logg\_std} & [cgs] & surface gravity, {\tt xgboost-distribution} standard deviation \\ 
10 & {\tt xgb\_met} &  & metallicity, {\tt xgboost} point estimate \\ 
11 & {\tt xgb\_met\_mean} &  & metallicity, {\tt xgboost-distribution} mean value \\ 
12 & {\tt xgb\_met\_std} &  & metallicity, {\tt xgboost-distribution} standard deviation \\ 
13 & {\tt xgb\_mass} & $M_{\odot}$ & stellar mass, {\tt xgboost} point estimate \\ 
14 & {\tt xgb\_mass\_mean} & $M_{\odot}$ & stellar mass, {\tt xgboost-distribution} mean value \\ 
15 & {\tt xgb\_mass\_std} & $M_{\odot}$ & stellar mass, {\tt xgboost-distribution} standard deviation \\ 
16 & {\tt dist} & mag & Distance estimate from the literature \\ 
17 & {\tt dist\_lower} & mag & 16th distance percentile from the literature \\ 
18 & {\tt dist\_upper} & mag & 84th distance percentile from the literature \\ 
18 & {\tt dist\_flag} & mag & Distance flag (0={\tt StarHorse} EDR3, 1=\citealt{Bailer-Jones2021} photogeo, =2 BJ2021 geo) \\ 
19 & {\tt bprp0} & mag & dereddened colour, derived with \href{https://github.com/fjaellet/gaia_edr3_photutils}{gaia\_edr3\_photutils} \\ 
20 & {\tt mg0} & mag & absolute magnitude, derived with \href{https://github.com/fjaellet/gaia_edr3_photutils}{gaia\_edr3\_photutils} \\ 
21 & {\tt xg} & kpc & Galactocentric Cartesian X coordinate, derived from {\tt dist} and assuming $X_0 = -8.2$ kpc \\ 
22 & {\tt yg} & kpc & Galactocentric Cartesian Y coordinate, derived from {\tt dist} and assuming $X_0 = -8.2$ kpc \\ 
23 & {\tt zg} & kpc & Galactocentric Cartesian Z coordinate, derived from {\tt dist} and assuming $Z_0 = 0$  \\ 
24 & {\tt rg} & kpc & Galactocentric planar distance, derived from {\tt XGal} and {\tt YGal} \\ 
25 & {\tt vxg} & km/s & Galactic Cartesian velocity in X direction \\ 
26 & {\tt vyg} & km/s & Galactic Cartesian velocity in Y direction \\ 
27 & {\tt vzg} & km/s & Galactic Cartesian velocity in Z direction \\ 
28 & {\tt vrg} & km/s & Galactic radial velocity \\ 
29 & {\tt vphig} & km/s & Galactic angular velocity\\ 
30 & {\tt xgb\_inputflag} &  & {\tt SHBoost} input flag \\ 
31 & {\tt xgb\_av\_outputflag} &  & $A_V$ output quality flag (=0 if {\tt xgbdist\_av\_std} $< 0.3$)\\ 
32 & {\tt xgb\_logteff\_outputflag} &  & $\log T_{\rm eff}$ output quality flag (=0 if {\tt xgbdist\_logteff\_std} $< 0.1$)\\ 
33 & {\tt xgb\_logg\_outputflag} &  & $\log g$ output quality flag (=0 if {\tt xgbdist\_logg\_std} $< 0.3$)\\ 
34 & {\tt xgb\_met\_outputflag} &  & [M/H] output quality flag (=0 if {\tt xgbdist\_met\_std} $< 0.3$)\\ 
35 & {\tt xgb\_mass\_outputflag} &  & Mass output quality flag (=0 if {\tt xgbdist\_mass\_std} / {\tt xgbdist\_mass\_mean} $< 0.3$)\\ 
\end{tabular}
\label{tab:datamodel1}
\end{table*}

\section{Empirical a posteriori metallicity calibration} \label{sec:feh_calib}

\begin{figure}\centering
        \includegraphics[width=0.49\textwidth]{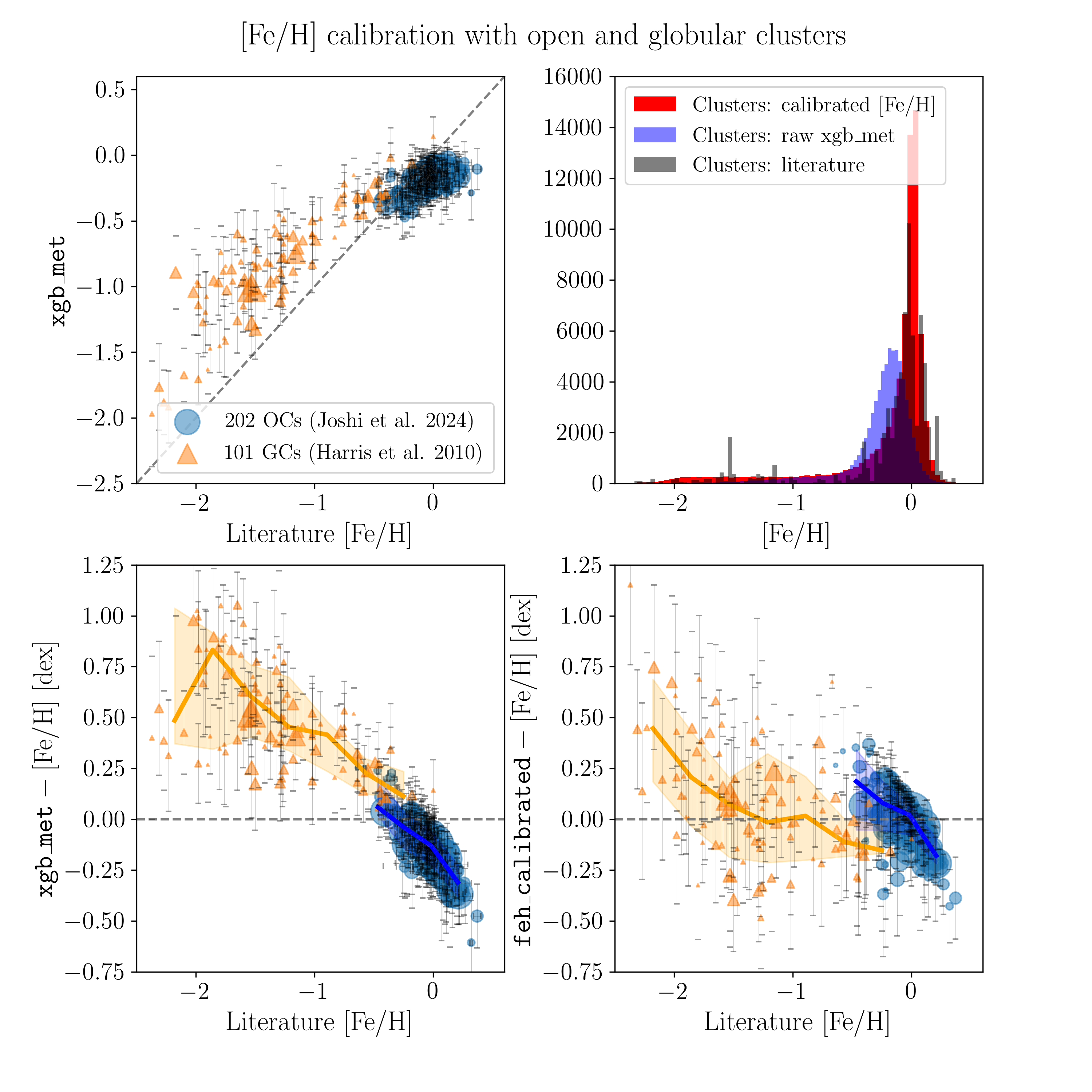}
        \caption{[Fe/H] calibration based on members of open and globular clusters. Top left panel: One-to-one comparison of the {\tt xgb\_met} values for open and globular cluster members with spectroscopic [Fe/H] measurements from the literature (using \citealt{Joshi2024} for open clusters and \citealt{Harris2010} for globular clusters). Second row: residuals between {\tt xgb\_met} and literature [Fe/H] (left panel) and {\tt feh\_calibrated} and literature [Fe/H], showing the improvement achieved by our proposed calibration. Top right panel: comparison of the three metallicity distributions (literature, {\tt xgb\_met}, and {\tt feh\_calibrated}) for the cluster sample.}
        \label{fig:feh_calib}
\end{figure}

\begin{figure}\centering
        \includegraphics[width=0.49\textwidth]{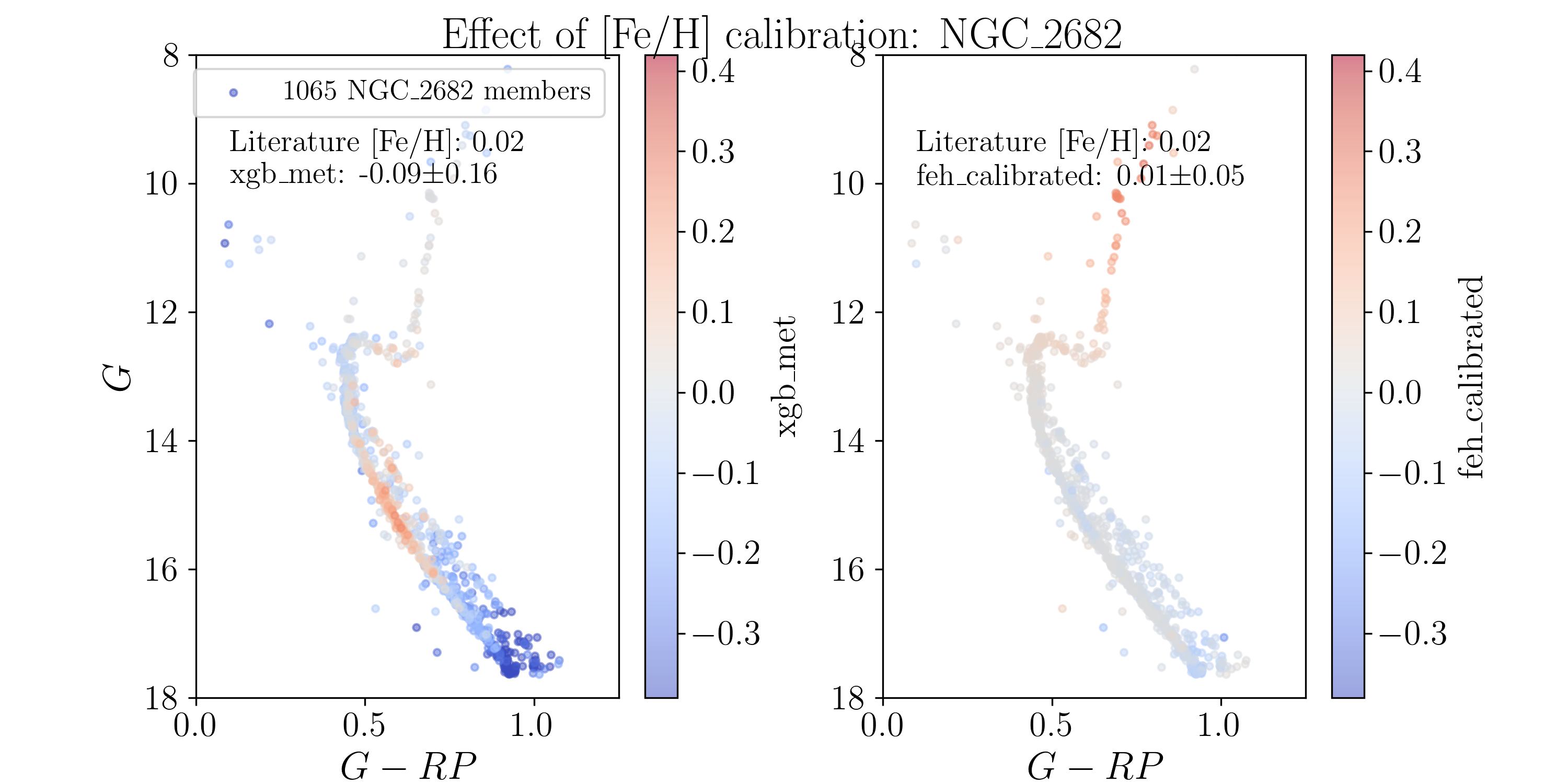}
        \includegraphics[width=0.49\textwidth]{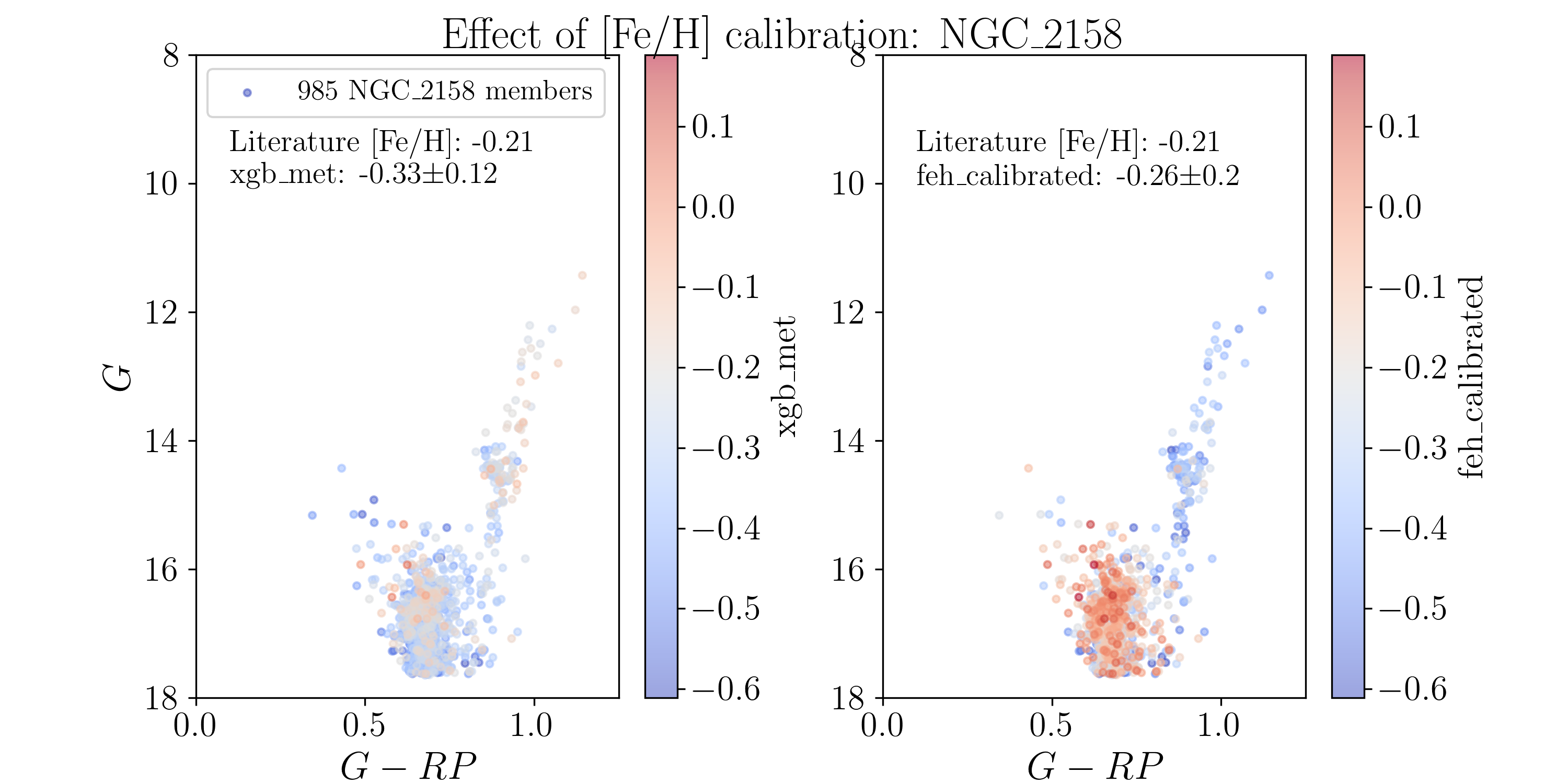}
        \includegraphics[width=0.49\textwidth]{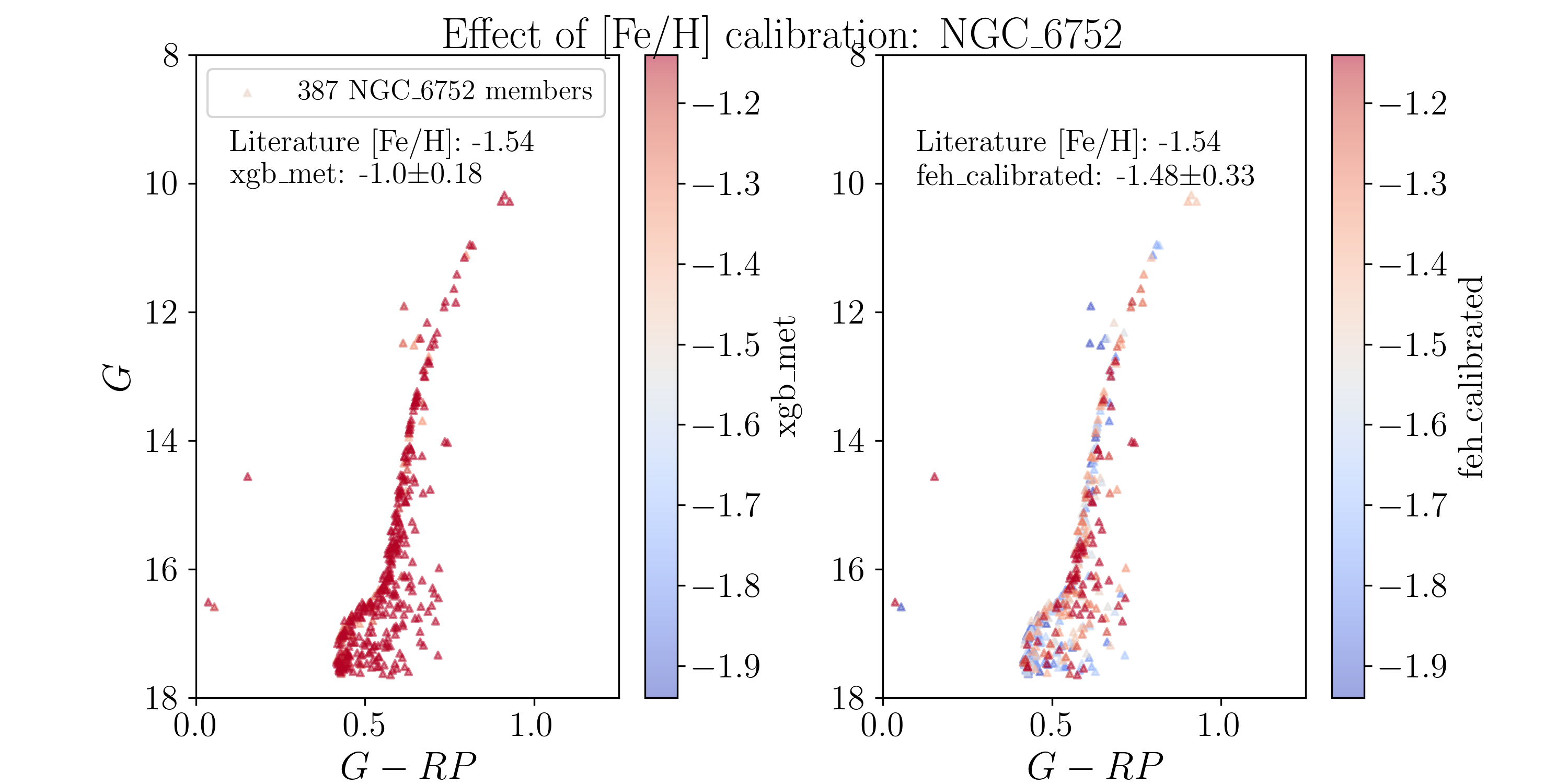}
        \caption{Effect of the [Fe/H] calibration for a few example open and globular clusters. Each panel shows the observed {\it Gaia} DR3 colour-magnitude diagrams ($G$ vs $G-G_{RP}$), colour-coded by either {\tt xgb\_met} (left panels) or {\tt feh\_calibrated} (right panels). The colour code limits are different in each row to highlight the differences with respect to the respective literature [Fe/H]. Top row: The solar-metallicity open cluster M67 (NGC 2682). Middle row: The metal-poor open cluster NGC 2158. Bottom row: Bulge globular cluster NGC 6752.}
        \label{fig:feh_calib2}
\end{figure}

When comparing the metallicity estimates obtained with {\tt xgboost} to a large sample of open and globular clusters (see Figs. \ref{fig:feh_calib} and \ref{fig:feh_calib2}), we find important systematic offsets that scale both with metallicity itself, but also with effective temperature and surface gravity. Since our raw {\tt xgb\_met} results correspond to [M/H] rather than [Fe/H], part of the trend with [Fe/H] (left panels of Fig. \ref{fig:feh_calib}) is partly expected due to $\alpha$-enhancement at low metallicities. In addition, overestimated metallicities for globular clusters have been found to be common even in spectroscopic surveys \citep{Soubiran2022}. However, we also find important residual trends with stellar parameters (see left panels of Fig. \ref{fig:feh_calib2}). We therefore decided to derive cluster-calibrated estimates of [Fe/H] for the full dataset, based on a third-order polynomial fit of $\{{\tt xgb\_met}, {\tt xgb\_logteff}, {\tt xgb\_logg}, {\tt phot\_g\_mean\_mag}\}$ to the residuals.

We cross-matched our {\it Gaia} DR3 XP sample with the recent star-cluster membership catalogue of \citet{Hunt2023}, which was in turn cross-matched with the [Fe/H] compilations of \citet{Joshi2024} and \citet{Harris2010}, resulting in a sample of 61660 stars in 202 open clusters and 9687 stars in globular clusters (blue and orange symbols in Fig. \ref{fig:feh_calib}, respectively). We find that two separate trends are discernible for stars with {\tt xgb\_met} $>-0.6$ and $\leq -0.6$, so these regimes were treated separately.
The results of our [Fe/H] calibration are shown in the right panels of Figs. \ref{fig:feh_calib} and \ref{fig:feh_calib2}, demonstrating that it is possible to (at least partly) calibrate out the large biases in {\tt xgb\_met} when compared to literature [Fe/H] measurements.
In the case of the open cluster M67 (a.k.a. NGC 2682; top panels of Fig. \ref{fig:feh_calib2}), for example, the calibrated [Fe/H] values (right panel) show a very small spread and bias with respect to the literature, while the {\tt xgb\_met} values do show significant trends as a function of position in the CMD. In other cases, however, the calibration does not improve the results significantly.

In this paper, however (in particular in Sects. \ref{sec:maps} and \ref{sec:vmp}), we use only the uncalibrated {\tt xgb\_met} values, keeping in mind that these tend to be significantly overestimated for low-metallicity objects, and slightly underestimated for super-solar metallicities (left panels of Fig. \ref{fig:feh_calib}). This is an effect that is commonly seen in machine-learning regression estimates of unbalanced datasets (e.g. \citealt{Guiglion2024}; see also Appendix \ref{sec:gui2024}).

\end{document}